\documentclass[12pt]{article}
\usepackage{amsmath,amssymb,mathrsfs,framed,diagrams}
\usepackage{amsmath,amssymb,mathrsfs,framed}
\newarrow {Corresponds} <{dash}{}{dash}>
\usepackage[colorlinks]{hyperref}

\usepackage{color}

\usepackage[all]{xy}

\newcommand{\bnu}{{\boldsymbol{\nu}}}
\newcommand{\bgamma}{{\boldsymbol{\gamma}}}

\newcommand{\bx}{{\bf x}}
\newcommand{\bn}{{\mathbf{n}}}
\newcommand{\bm}{{\mathbf{m}}}
\newcommand{\de}{{\mathrm{d}}}

\newcommand{\rem}[1]{}

\makeatletter

\@addtoreset{figure}{section}
\def\thefigure{\thesection.\@arabic\c@figure}
\def\fps@figure{h, t}
\@addtoreset{table}{bsection}
\def\thetable{\thesection.\@arabic\c@table}
\def\fps@table{h, t}
\@addtoreset{equation}{section}

\makeatother

\newcommand{\todo}[1]{\vspace{5 mm}\par \noindent
\framebox{\begin{minipage}[c]{0.95 \textwidth}
\tt #1 \end{minipage}}\vspace{5 mm}\par}

\begin{document}

\newtheorem{theorem}{Theorem}[section]
\newtheorem{definition}[theorem]{Definition}
\newtheorem{lemma}[theorem]{Lemma}
\newtheorem{remark}[theorem]{Remark}
\newtheorem{proposition}[theorem]{Proposition}
\newtheorem{corollary}[theorem]{Corollary}
\newtheorem{example}[theorem]{Example}

\def\below#1#2{\mathrel{\mathop{#1}\limits_{#2}}}



\title{Equivalent theories of liquid crystal dynamics}
\author{Fran\c{c}ois Gay-Balmaz$^{1}$, Tudor S. Ratiu$^{2}$, Cesare Tronci$^{3}$}
\addtocounter{footnote}{1} 
\footnotetext{CNRS / Laboratoire de 
M\'et\'eorologie Dynamique, \'Ecole Normale Sup\'erieure, 
Paris, France. \textcolor{black}{Partially supported by a ÒProjet Incitatif de RechercheÓ contract from the Ecole Normale
Sup\'erieure de Paris.}
\texttt{gaybalma@lmd.ens.fr}
\addtocounter{footnote}{1} }
\footnotetext{Section de Math\'ematiques and Bernoulli
Center, \'Ecole Polytechnique F\'ed\'erale de
Lausanne, CH--1015 Lausanne. Switzerland. Partially supported by Swiss NSF grant 200021-140238 and by the government grant of the Russian Federation for support of research projects implemented by leading scientists, Lomonosov Moscow State University under  agreement No. 11.G34.31.0054. 
\texttt{tudor.ratiu@epfl.ch}
\addtocounter{footnote}{1} }
\footnotetext{Department of Mathematics, University of
Surrey,  Guildford, Surrey, GU2 7XH, United Kingdom. 
\texttt{c.tronci@surrey.ac.uk}
\addtocounter{footnote}{1} }

\date{ }
\maketitle

\makeatother
\maketitle




\begin{abstract}
There are two competing descriptions of nematic liquid 
crystal dynamics: the Ericksen-Leslie director theory
and the Eringen micropolar approach. Up to this day,
these two descriptions have remained distinct in spite
of several attempts to show that the micropolar theory includes the director theory. In this paper we show
that this is the case by using symmetry 
reduction techniques and introducing a new system
that is equivalent to the Ericksen-Leslie
equations and includes disclination dynamics.
The resulting equations of motion are verified to be completely equivalent, although one of the two different reductions offers the possibility of accounting for orientational defects. 
After applying these two approaches to the ordered micropolar theory of Lhuiller and Rey, all the results are eventually extended to flowing complex fluids, such as nematic liquid crystals.
\end{abstract}

\tableofcontents


\section{Introduction}

The Ericksen-Leslie (EL) equations for the dynamics of nematic liquid crystals are widely accepted and have been experimentally verified \cite{deGennes1971,Chandra1992,ReyDenn}. However, when  orientational defects (\emph{disclinations}) are present in the system, this model requires further development
in order to provide a reliable description. For example, in the presence of defects, the liquid crystal molecules may undergo phase transitions, e.g., from uniaxial to biaxial, and the director field $\bn$ in the Ericksen-Leslie equations is no longer an appropriate order parameter variable.

Among the various descriptions that incorporate defect dynamics, the micropolar theory developed by Eringen \cite{Eringen1993, Eringen1997} provides a general description of the motion of microfluids, including liquid crystals. Indeed, besides incorporating molecular shape {\color{black}effects} into a microinertia tensor $j$, the Eringen model encodes disclination dynamics in the so called \emph{wryness tensor} $\gamma$, which is expressed in terms of $(\nabla\mathbf{n})\times\bn$ when defects are absent \cite{Eringen1997}.

Nematic liquid crystals comprise a familiar example of microfluids. However, despite several
attempts, the EL description has not yet been derived from Eringen's micropolar theory. For example, the relation $\bgamma=(\nabla\mathbf{n})\times\bn$ proposed by Eringen 
\cite[formula (11.2)]{Eringen1997} fails to return the correct EL equations \cite{Le1979} as shown in
\cite[Theorem 8.11]{GBRa2009} by two different methods (symmetry considerations and a direct computation). Thus
it is not completely clear how $\bgamma$ may be expressed in terms of the director $\bn$. However, we 
point out that in \cite{Eringen1993}, Eringen himself 
realized that the definition of $\boldsymbol{\gamma}$ is not 
determined uniquely in terms of $\mathbf{n}$ and, therefore, the relation $\boldsymbol{\gamma} = \nabla\mathbf{n}\times\mathbf{n}$ cannot be used to formulate a consistent theory, 
although ``the non-uniqueness of $\boldsymbol{\gamma}$ does
not affect the free energy'' (see 
\cite[page 612]{Eringen1993}).

Recent new understanding of defect dynamics has been obtained from reduction theory \cite{Ho2002,GBRa2009}, which underlies the gauge-theory approach  \cite{Volovick1980}. {\color{black}This theory} applies to very general systems since it incorporates {\color{black}disclination} dynamics in different contexts, such as frustrated spin glasses 
\cite{HoKu1988,Volovick1980}, for example. In this setting, one is naturally led to consider the wryness tensor $\bgamma$ as the magnetic vector potential of a Yang-Mills field (or, equivalently, a \emph{connection 
one-form}) taking values in the Lie algebra $\mathfrak{so}(3)$ of antisymmetric $3 \times 3$ matrices (usually identified with vectors in $\Bbb{R}^3$) of the rotation group $SO(3)$. The quantity $\boldsymbol\gamma$ is also known as `spatial rotational strain' \cite{Ho2002} and it expresses the amount by which a specified director field rotates under an infinitesimal displacement. Due to its tensorial nature, the gauge potential $\bgamma$ may be conveniently expressed in terms of an appropriate basis as
\begin{equation}
\label{tensor_gamma}
\bgamma=\bgamma_i\,\de x^i
= \boldsymbol{\gamma}^a \mathbf{e}_a 
=\gamma_i^a\,\mathbf{e}_a\,\de x^i 
\end{equation}
where $\{\mathbf{e}_a\}$ is a fixed basis of $\Bbb{R}^3\simeq\mathfrak{so}(3)$. Then, its corresponding magnetic vector field is given componentwise by
\begin{equation}\label{Magnetic-Field}
\boldsymbol{B}^i=\epsilon^{ijk}\!\left(\partial_j\bgamma_k+\bgamma_j\times\bgamma_k\right),
\end{equation}
where we sum over repeated indices and we have used the equivalence between two-forms and vector fields on physical space (see \S\ref{Sec:micropolar} for the coordinate-free definition).
In the gauge-theory approach developed in \cite{Volovick1980}, the absence of disclinations is given by a vanishing magnetic field $\boldsymbol{B}$, rather than by a vanishing potential $\bgamma$. Thus, the presence of $\bgamma$ in a mathematical model must be compatible with EL dynamics, as long as $\boldsymbol{B}=0$. In the context of reduction theory, one recognizes that a vanishing magnetic field $\boldsymbol{B}=0$ simply amounts to the homogeneous initial condition $\bgamma_0=0$ \cite{Volovick1980}. If the latter condition is not satisfied, then the gauge-theory model would extend the EL formulation to incorporate non-trivial disclination dynamics. 

On the other hand, Eringen's micropolar theory does not seem to possess a gauge-theory formulation, since the wryness tensor $(\nabla\bn)\times\bn$, as defined by Eringen, does not transform as a magnetic potential under $SO(3)$ gauge transformations; see  \cite[Lemma 8.10]{GBRa2009}. Nevertheless, Eringen's theory still shares many analogies with gauge-theory models and the coexistence of the wryness and microinertia tensors in the dynamics provides an interesting opportunity to address for the shape evolution of the molecules interacting with disclination lines.

These considerations motivate the present work, which uses  Euler-Poincar\'e variational methods to provide a unifying framework for incorporating defect dynamics in continuum systems with broken internal symmetry (e.g., liquid crystals) and shows
that Eringen's micropolar theory includes Ericksen-Leslie dynamics. This is done upon noticing that taking the gradient of the relation
\[
\bn(\mathbf{x},t)=\chi(\mathbf{x},t)\,\mathbf{e}_3,
\]
that relates director dynamics to the dynamics of the rotation matrix $\chi(\mathbf{x},t)\in SO(3)$ in EL theory, immediately leads to 
\[
\nabla\bn=(\nabla\chi)\mathbf{e}_3
=(\nabla\chi)\chi^{-1\,}\bn\,.
\]
Here $\mathbf{e}_3:=(0,0,1)$. Then, one observes that the new variable 
\begin{equation}\label{bgamma}
\widehat{\bgamma}:=-(\nabla\chi)\chi^{-1}
\end{equation}
is precisely a connection one form taking values in $\mathfrak{so}(3)$ \cite{Ho2002,GBRa2009}. It is straightforward to see that analogous relations hold independently of the order parameter space. Then, upon using the isomorphism $\mathfrak{so}(3)\simeq\Bbb{R}^3$ given by the (inverse of the) hat map 
\begin{equation}
\label{hatmap}
a^i=-\epsilon^{ijk\;}\widehat{a}_{jk}\,,
\qquad
\forall\,\widehat{a}\in\mathfrak{so}(3)
\,, 
\end{equation}
one can simply replace the relation 
\[
\nabla\bn=\bn\times\bgamma
\]
 into the EL equations to account for the potential 
$\bgamma$ as an extra dynamical variable. Notice that, although the latter relation is satisfied by the choice $\bgamma=(\nabla\bn)\times\bn$, this expression is only defined up to a component  parallel to $\bn$. Thus, $\bgamma$ cannot be entirely expressed in terms of the director $\bn$ and it needs to be specified by all three columns of the matrix $\chi(\mathbf{x},t)$. 

The second key observation is that a different symmetry reduction
of the same material Lagrangian yields a new set of
equations for nematodynamics. We show that these
are completely equivalent to the original Ericksen-Leslie
equations. However, this new system allows for the
description of disclinations, something that the Ericksen-Leslie equations could not handle, as discussed
above.

As we shall see, all the above considerations hold regardless of the background fluid motion and they are a particular feature of the micro-order. Thus, we shall mainly confine our treatment to motion-less liquid crystal continua in order to emphasize the high points of the discussion. The extension to flowing fluid systems will be presented briefly at the end of this paper.

\begin{remark}[Dissipative vs. conservative dynamics] {\rm Notice that this paper neglects dissipative terms in order to focus on inertial effects of liquid crystal dynamics. A possible strategy for including dissipation within the same treatment is found in \cite{GBRaTr2012}. As will
become clear from our treatment, the relationships among
various models of nematodynamics established in this paper, remain unchanged when dissipation is added (e.g., by Rayleigh's method \cite{BlBrMaRa,GBRaTr2012}, double bracket terms \cite{BlBrMaRa,HoPuTr}, etc).
\quad $\blacklozenge$}
\end{remark}

\rem{ 
\todo{Shouldn't we say something about the confusion
in Eringen's papers here? Should we mention the computation in the Advances paper where we checked both
by hand and from symmetry that the formula for $\boldsymbol{\gamma}$ is false?\\
F: Cesare, can you write a text about that? In \cite{GBRa2009} there were two observations:\\
- first (Lemma 8.10) if $\mathbf{n}$ evolves as a director field, then $\bgamma=\nabla\bn\times\bn$ does not evolve as a connection (this already shows that this is not a good definition).\\
- Second (Theorem 8.11): if we take the Ericksen-Leslie equations and define $j=J(\mathbf{I}-\mathbf{n}\otimes\mathbf{n})$ and $\bgamma=\nabla\bn\times\bn$ and suppose that we can write $F(\mathbf{n},\nabla\mathbf{n})=\Psi(j,\bgamma)$ (which is the case for the Frank energy), then the resulting equations of motion are close but different from Eringen equations.\\
}
} 
\paragraph{Plan of the paper.} This paper starts (Section 2) by showing how reduction theory can be applied to Ericksen-Leslie nematodynamics in two different fashions, thereby producing two different sets of equations of motion. The resulting dynamical systems are, however, completely equivalent. In Section 3, these two equivalent reduction methods are then formulated in a general context, for an arbitrary order parameter space. Momentum map properties are presented in detail for the two constructions, which are then specified to micropolar continua. In Section 4, Eringen's theory of micropolar media is shown to comprise Ericksen-Leslie nematodynamics. This requires a specified choice of the micropolar free energy, which in turn reduces to the Frank energy under the assumption of uniaxial molecules. While Section 5 deals with the Lhuiller-Rey theory of ordered micropolar continua \cite{LR}, Section 6 extends all the results to liquid crystal flows, thereby showing how the hydrodynamic Ericksen-Leslie equations possess a micropolar formulation.

\section{Two equivalent reductions for nematic systems}

This section develops the guiding example of this  paper, i.e., the dynamics of nematic media. In particular, this section shows how the reduction producing EL nematodynamics is accompanied by an equivalent reduction procedure that naturally incorporates the connection $\gamma=-(\nabla\chi)\chi^{-1}$ as an extra dynamical variable. The latter construction will be presented after the following review of the reduction underlying EL dynamics.

\subsection{Notation}
We regard the director field as a smooth map $\mathbf{n}:
\mathcal{D}\rightarrow S^2$. In more generality, it is convenient to introduce the notation
\[
\mathcal{F(D},M):=\{f\ |\ f:\mathcal{D}\to M\}
\,,
\]
which defines the set of all smooth mappings $\mathcal{D}\to M$, where $M$ is some differentiable manifold. In this notation, we have $\bn\in\mathcal{F(D},S^2)$ and $\chi\in\mathcal{F(D},SO(3))$. In particular, $\mathcal{F(D},SO(3))$ is referred to as the \emph{gauge group}. 

In a similar fashion, we denote by $\Omega^k(\mathcal{D}, V)$,
the space of exterior differential $k$-forms taking values in
the vector space $V$. Analogously, 
$\mathfrak{X}(\mathcal{D}, V)$ denotes the space of vector
fields on $\mathcal{D}$ taking values in $V$, i.e., contravariant $V$-valued one-tensors.

Conforming with standard notation used in both elasticity
theory and liquid crystals, in this paper $\nabla$ denotes 
the derivative (or tangent) of a map between two 
manifolds or the exterior derivative if the target manifold 
is a vector space. Thus, for example, if $n \in 
\mathcal{F}(\mathcal{D}, M)$, then  $\nabla n:= Tn: 
T \mathcal{D} \rightarrow TM$, and if $M =\mathbb{R}^3$, then 
$\nabla \mathbf{n}:= \mathbf{d} \mathbf{n}: T \mathcal{D} \rightarrow
\mathbb{R}^3$ is the usual $\mathbb{R}^3$-valued exterior
differential of a function, i.e., a $\mathbb{R}^3$-valued
one-form on $\mathcal{D}$. If $\mathcal{O}$ is a Lie group
and $\chi\in \mathcal{F}(\mathcal{D},\mathcal{O})$, then
$\nabla \chi: = T \chi: T \mathcal{D} \rightarrow T 
\mathcal{O}$. As opposed to standard notation in Riemannian
geometry, if $p \in \mathcal{F}(\mathcal{D},\mathbb{R})$,
then $\nabla p : = \mathbf{d} p$, the exterior derivative
of the function $p $, so, with these conventions, $\nabla p$
is an exact one-form on $\mathcal{D}$.

In this paper, $\mathcal{D} \subset \mathbb{R}^3$ has 
non-empty interior and if it has a boundary, then it is smooth. However, all results and their proofs remain
unchanged for an arbitrary smooth oriented manifold 
$\mathcal{D}$, possibly with smooth boundary, with volume 
form $\mu$; one needs only change $\nabla$ to the appropriate obvious 
derivative in the calculus on manifolds and assume appropriate boundary conditions.

{\color{black}
\subsection{Background on the Ericksen-Leslie and Eringen models}
The director field of a nematic medium takes values in the space of  unsigned unit vectors $\bn(\bx,t)\in S^2$, with $\bn\sim-\bn$. In the physics literature, it is customary to work simply with unit vectors in physical space $\Bbb{R}^3$, by making sure that all relations are invariant under reflections. This avoids many complications that may arise from working on the real projective plane. 

The dynamics of the director field is typically governed by the Ericksen-Leslie equations
\begin{equation}\label{ELeslie}
J\frac{\partial^2\mathbf{n} }{\partial t^2}-\left(\mathbf{n\cdot h}+J\,\mathbf{n}\cdot
\frac{\partial^2\mathbf{n} }{\partial t^2}\right)\mathbf{n}
+\mathbf{h}=0
\,.
\end{equation}
Here,  $J$ is the \emph{microinertia constant}, while the \emph{molecular field} 
\begin{equation}
\mathbf{h}:=\frac{\partial  F}{\partial  \mathbf{n}}-
\frac{\partial}{\partial x^i}\frac{\partial  F}{\partial  (\partial_{x^i}\bn)}
\label{molecularfield1}
\end{equation}
is expressed in terms of the Frank energy $F$:
 \begin{align}
F(\mathbf{n},{\nabla}\mathbf{n})&:=K_2\underbrace{(\mathbf{n}\cdot\operatorname{curl}\mathbf{n})}_{\textit{chirality}}+\frac{1}{2}K_{11}\underbrace{(\operatorname{div}\mathbf{n})^2}_{\textit{splay}}+\frac{1}{2}K_{22}\underbrace{(\mathbf{n}\cdot\operatorname{curl}\mathbf{n})^2}_{\textit{twist}}
+\frac{1}{2}K_{33}\underbrace{\|\mathbf{n}\times\operatorname{curl}\mathbf{n}\|^2}_{\textit{bend}},
\label{standard_energy}
\end{align}
where each term possesses a precise physical meaning, as indicated above. (Here, $K_2\neq 0$ for cholesterics and $K_2=0$ for nematics). 
The free energy can also contain additional terms due to 
external electromagnetic fields. 

It is easy to verify that equation \eqref{ELeslie} is an 
Euler-Lagrange equation on the tangent bundle $TS^2$, as it 
arises from the Lagrangian
 \begin{equation}\label{lagr-ELeslie}
{\sf L}(\mathbf{n},\partial_t{\mathbf{n}})=
\frac{J}{2}\int_\mathcal{D}\left\|\partial_t{\mathbf{n}}
\right\|^2\,\de^3 \mathbf{x}-
\int_\mathcal{D}F(\mathbf{n},
\nabla\mathbf{n})\,\de^3\mathbf{x}
\,,
\end{equation}
where $\mathcal{D}\subset\mathbb{R}^3$ is the spatial domain 
occupied by the nematic medium. Notice that the Euler-Lagrange equations arising from the above Lagrangian will require using covariant derivatives on the unit sphere (as shown in \cite{GBTr2010}). In what follows, the volume element $\de^3\bx$ will be replaced by the measure $\mu$. Since 
$\mathbf{n}(\mathbf{x},t)\in S^2$, then it is convenient to encode its rotational dynamics into an orthogonal matrix 
$\chi(\mathbf{x},t)\in SO(3)$ so that
\[
\mathbf{n}(\mathbf{x},t)=\chi(\mathbf{x},t)\mathbf{n}_0
\]
where $\mathbf{n}_0$ is the initial condition on the director. In this paper, we shall consider the particular case in which $\mathbf{n}_0=\mathbf{e}_3=(0,0,1)$. Then, the director dynamics can be expressed in terms of the evolution of $\chi$. 

After introducing the rotation matrix $\chi$, Eringen's theory uses the relation $\nabla\bn=-{\bgamma}\times\bn$, where $\widehat{\bgamma}$ is given by \eqref{bgamma}. Then, the dynamics can be expressed in terms of $\bn$ and $\bgamma$. The relation $\nabla\bn=-{\bgamma}\times\bn$ has the algebraic solution $\bgamma=(\nabla\bn)\times\bn$, which is used in Eringen's work \cite[formula (11.2)]{Eringen1997}. However, a careful analysis shows that, since it is not satisfied at all times, this algebraic solution yields an erroneous dynamical theory. 
The next sections analyze this situation in more detail from the point of view of Euler-Poincar\'e reduction by symmetry. As we shall see, the relation $\nabla\bn=-{\bgamma}\times\bn$ yields a consistent theory of nematic dynamics, without the identification $\bgamma=(\nabla\bn)\times\bn$.
}

\subsection{Reduction for the Ericksen-Leslie equations}\label{sec:ErLe}
The reduction process producing the EL equations has been widely explained in \cite{Ho2002,GBRa2009}. This process starts by identifying the configuration space of a nematic continuum with the space $\mathcal{F}(\mathcal{D},SO(3))$ of $SO(3)$-valued  functions on the domain $\mathcal{D}\subset\Bbb{R}^3$. Then, one makes use of the Lagrangian \cite{Ho2002,GBRa2009}
\begin{equation}\label{UnreducedLagrangian}
\mathcal{L} ( \chi , \dot \chi )=\frac{1}{2} J\int_{ \mathcal{D} } \|\dot{\chi}{\color{black}\bn_0}\|^2  \mu -\int_{ \mathcal{D} }F( \chi \mathbf{n} _0 , {\nabla} ( \chi \mathbf{n} _0 )) \mu\, ,
\end{equation}
where, as usually, $\mathbf{n}_0={\color{black}\mathbf{e}_3}$ (although it can be an arbitrary director field $\bn_0(\mathbf{x})$)  and $J$ is the microinertia constant. 
\rem{ 
Here the free energy is given by the Frank expression
\begin{align}\label{standard_energy}
F(\mathbf{n},{\nabla}\mathbf{n})&=K_2\underbrace{(\mathbf{n}\cdot\operatorname{curl}\mathbf{n})}_{\textbf{chirality}}+\frac{1}{2}K_{11}\underbrace{(\operatorname{div}\mathbf{n})^2}_{\textbf{splay}}+\frac{1}{2}K_{22}\underbrace{(\mathbf{n}\cdot\operatorname{curl}\mathbf{n})^2}_{\textbf{twist}}
\nonumber \\
& \qquad
+\frac{1}{2}K_{33}\underbrace{\|\mathbf{n}\times\operatorname{curl}\mathbf{n}\|^2}_{\textbf{bend}},
\end{align}
(here, $K_2\neq 0$ for cholesterics and $K_2=0$ for nematics). 
The free energy can also contain additional terms due to external
electromagnetic fields.
} 
{\color{black}In order} to apply the Euler-Poincar\'e theory for systems with broken symmetry (see \cite{HoMaRa1998,GBTr2010}), we {\color{black} define an extended Lagrangian $L$ by writing} $\mathcal{L}(\chi,\dot\chi)=:L(\chi,\dot\chi,\mathbf{n}_0)$, where $L:T\mathcal{F(D,}SO(3))\times\mathcal{F}(\mathcal{D},S^2) \rightarrow \mathbb{R}$ {\color{black} is a functional on the extended domain $T\mathcal{F(D,}SO(3))\times\mathcal{F}(\mathcal{D},S^2)$. Observe} that $L$ is invariant under the {\color{black}following right action of the gauge group $\mathcal{F(D,}SO(3))$ on the extended space $T\mathcal{F(D,}SO(3))\times\mathcal{F}(\mathcal{D},S^2)$:}
\[
( \chi ,  \mathbf{n} _0) \mapsto \left( \chi \psi , \psi ^{-1}  \mathbf{n} _0\right)
\,,
\]
where $\psi\in\mathcal{F}(\mathcal{D},SO(3))$. This invariance property yields the reduced Euler-Poincar\'e Lagrangian
\begin{equation}
\ell_1( \boldsymbol{\nu},  \mathbf{n} )= \frac{1}{2} J\int_{ \mathcal{D} } \|\boldsymbol{\nu}{\color{black}\times\bn}\|^2\mu -\int_{ \mathcal{D} }F ( \mathbf{n} , {\nabla} \mathbf{n} ) \mu
\,,
\label{ELlagrangian1}
\end{equation}
where $\widehat{\bnu}=\dot\chi \chi^{-1} $ and $\bn=\chi\bn_0$, where $\boldsymbol{\nu}\in\mathbb{R}^3\mapsto \widehat{\boldsymbol{\nu}}\in\mathfrak{so}(3)$ denotes the usual Lie algebra isomorphism defined by $\widehat{\boldsymbol{\nu}}_{ab}=-\epsilon_{abc}\boldsymbol{\nu}_c$. We thus obtain the following equations \cite{Ho2002,GBRa2009}
\begin{equation}\label{EL}
\left\{
\begin{array}{l}
\vspace{0.2cm}\displaystyle\frac{\partial}{\partial t}\frac{\delta \ell _1 }{\delta \boldsymbol{\nu}}=\boldsymbol{\nu}\times \frac{\delta \ell _1 }{\delta \boldsymbol{\nu}} +\mathbf{n} \times \frac{\delta \ell _1 }{\delta \mathbf{n} }\\
\displaystyle \partial_ t \mathbf{n} + \mathbf{n} \times \boldsymbol{\nu }=0
\end{array}\right.
\end{equation}
by applying the usual Euler-Poincar\'e variational principle 
\begin{equation}\label{HamPrinc1}
\delta\int^{t_1}_{t_0}\ell_1(\bnu,\bn)\,\de t=0\,,
\end{equation}
subject to the variations $\delta\bnu=\partial_t{\boldsymbol\eta}+\bnu\times\boldsymbol\eta$ and $\delta\bn=\boldsymbol\eta\times\bn$ for arbitrary 
$\boldsymbol{\eta} \in{\color{black}\Bbb{R}^3\simeq \mathfrak{so}(3)}$ satisfying
$\boldsymbol{\eta}(t_0)= \boldsymbol{\eta}(t_1)=0$. {\color{black}More precisely, these variational relations arise from the definitions $\widehat{\bnu}=\dot\chi \chi^{-1}$ and $\mathbf{n}=\chi\mathbf{e}_3$, respectively, upon defining  $\widehat{\boldsymbol\eta}=(\delta\chi)\chi^{-1}$.

At this point, upon computing the variational derivatives 
\begin{align*}
&\frac{\delta \ell_1}{\delta\bnu}=-J\,\mathbf{n}\times(\mathbf{n}\times \boldsymbol{\nu})=J\,\mathbf{n}\times\partial_t\bn
\,,\\
&\frac{\delta \ell_1}{\delta\bn}=-J\,\boldsymbol{\nu}\times(\boldsymbol{\nu}\times\mathbf{n})-\mathbf{h}=-J\,\boldsymbol{\nu}\times\partial_t\bn-\mathbf{h},
\end{align*}
equations \eqref{EL} become
\begin{equation}
\label{EP_form_EL}
\left\{
\begin{array}{l}
\vspace{0.2cm}J\,\displaystyle{\partial_t}\big(\mathbf{n}\times(\mathbf{n}\times \boldsymbol{\nu})\big)=J\,\boldsymbol{\nu}\times \big(\mathbf{n}\times(\mathbf{n}\times \boldsymbol{\nu})\big) +\mathbf{n} \times \big(J\,\boldsymbol{\nu}\times(\boldsymbol{\nu}\times\mathbf{n})+\mathbf{h}\big) \\
\displaystyle \partial_ t \mathbf{n} + \mathbf{n} \times \boldsymbol{\nu }=0
\end{array}\right.
\end{equation}
which reduce to
\[
J\,\displaystyle{\partial_t}\big(\mathbf{n}\times\partial_ t \mathbf{n}\big)=J\,\boldsymbol{\nu}\times \big(\mathbf{n}\times\partial_ t \mathbf{n}\big) 
-\mathbf{n} \times \big(J\,\bnu\times\partial_ t \mathbf{n}+\mathbf{h}\big) = - \mathbf{n} \times \mathbf{h}\,,
\]
by he Jacobi identity. Thus we get the Ericksen-Leslie equations \eqref{ELeslie} with molecular field $\bf h$ given by \eqref{molecularfield1}.}

\rem{ 
 More explicitly, upon denoting $\mathbf{h}=-{\delta \ell_1}/{\delta \mathbf{n}}$, one has
\begin{equation}\label{EL-bis}
\left\{
\begin{array}{l}
\vspace{0.2cm}\displaystyle J\partial_t \boldsymbol{\nu}=  \mathbf{h}\times\mathbf{n}  \\
\displaystyle \partial_ t \mathbf{n} + \mathbf{n} \times \boldsymbol{\nu }=0,
\end{array}\right.
\end{equation}
which produce the EL equations of nematodynamics \cite{Le1979} (upon setting ${\bnu\cdot\bn=0}$)
\begin{equation}\label{ELeslie}
J\frac{d^2\mathbf{n} }{dt^2}-2\left(\mathbf{n\cdot h}+J\,\mathbf{n}\cdot\frac{d^2\mathbf{n} }{dt^2}\right)\mathbf{n}+\mathbf{h}=0\,.
\end{equation}
}    

Notice that we have {\color{black}$\bn\cdot\delta\ell_1/\delta\bnu=0$, which means that there is no angular momentum ${\color{black}\delta\ell_1/\delta\bnu}$ in the direction of the director $\bn$. This is interpreted in terms of the rod-like nature of  uniaxial molecules}, as explained in \cite{GBTr2010}. However, when disclination lines are present, molecules may change their shape (e.g., from uniaxial to biaxial) and the projection $\color{black}\bn\cdot\delta\ell_1/\delta\bnu$ cannot be a constant. The next section considers an alternative reduction for nematic systems, which can be naturally extended to account for disclination dynamics.
{\color{black}\noindent
\begin{remark}[Kinetic energy]{\rm 
Notice that one can replace the first term in the Lagrangian \eqref{UnreducedLagrangian} by the expression
\[
\frac{1}{2} J\int_{ \mathcal{D} } \|\dot{\chi}\|^2  \mu
\]
in which case, the same invariance holds and the first term in the reduced Lagrangian $\ell_1$ reads
\[
\frac{1}{2} J\int_{ \mathcal{D} } \|\boldsymbol{\nu} \|^2  \mu.
\]
The Ericksen-Leslie equations are recovered by imposing the constraint $ \boldsymbol{\nu}_0 \cdot \mathbf{n}_0  =0$ on the initial conditions. This constraint is preserved by the dynamics, see Lemma 8.3 \makebox{in \cite{GBRa2009}.}
\quad $\blacklozenge$}
\end{remark}
}\noindent

\rem{ 
Ericksne-Leslie is consistent with Lhuillier-Rey (no disclination dynamics)

Problems of compatibility between Lhuillier-Rey and Eringen.

We construct a system which encompass all the properties of all the other models: variables are $\nu , n, j, \gamma $. Show that $\nu\cdot n=0$ is not preserved: then uniaxial particles change their shape. Thus, nematics can turn into biaxials.

The same set of equations arise from an alternative reduction of the Lhuillier-Rey system. Although now we allow for nonzero curvature.
} 

\subsection{Alternative reduction for nematic systems}
While the EL equations are well known and widely accepted as a reliable model, this section presents an alternative set of equations that are completely equivalent to EL dynamics. These new equations have the advantage that 
they can be easily extended to consider non-trivial disclination effects, as we shall see later.

The starting point is the same unreduced Lagrangian \eqref{UnreducedLagrangian} producing EL dynamics. 
We notice that, if $\bn_0$ is an arbitrary constant director, the Lagrangian $\mathcal{L}$ in \eqref{UnreducedLagrangian} possesses the alternative invariance property
\[
\mathcal{L}(\chi,\dot\chi)=L(\dot\chi\chi^{-1},\chi\bn_0,-({\nabla}\chi)\,\chi^{-1})
\]
where ${\nabla}$ denotes the usual differentiation operator and $\gamma:=-({\nabla}\chi)\,\chi^{-1} \in \Omega ^1( \mathcal{D},\mathfrak{so}(3))$ 
is a new dynamical variable. More precisely, if $ \mathbf{n} _0 $ is a constant vector field, then we can rewrite the Lagrangian $ \mathcal{L} $ in \eqref{UnreducedLagrangian} as
\begin{equation}\label{SecondSymmetry}
\mathcal{L} ( \chi , \dot \chi )=\frac{1}{2}J\int_{ \mathcal{D} } \|\dot{\chi} {\mathbf{n}_0}\|^2 \mu -\int_{ \mathcal{D} }F\left( \chi \mathbf{n} _0 , (({\nabla} \chi) 
\chi ^{-1} )( \chi  \mathbf{n} _0 ) \right) \mu
\end{equation}
and consider $ \mathcal{L} $ as coming from a Lagrangian $L=L( \chi , \dot \chi ,\mathbf{n} _0, \gamma _0 )$ defined on $T \mathcal{F} ( \mathcal{D} , SO(3)) \times\mathcal{F} ( \mathcal{D} , \mathbb{R}  ^3 ) \times \Omega ^1 ( \mathcal{D} , \mathfrak{so}(3) )  $, which is invariant under the right action
\[
( \chi ,  \mathbf{n} _0, \gamma _0  ) \mapsto \left( \chi \psi , \psi ^{-1}  \mathbf{n} _0 , \psi ^{-1} \gamma _0 \psi + \psi ^{-1} {\nabla} \psi \right).
\]
In the present case, the initial value of $ \gamma $ is zero, that is, we have $\gamma_0=0$ so that
\[
\mathcal{L}(\chi,\dot\chi) =L(\chi,\dot\chi,\mathbf{n} _0 ,0)
\,.
\]
Then, the reduced Lagrangian corresponding to 
\eqref{SecondSymmetry} takes the form
\begin{align}
L(\dot\chi\chi^{-1},\chi\bn_0,-({\nabla}\chi)\,\chi^{-1})
&=\frac{1}{2} J\int_{ \mathcal{D} } {\color{black}\left\|\bnu\times\bn\right\|^2}  \mu -\int_{ \mathcal{D} }F( \mathbf{n} , -\bgamma\times \mathbf{n}  ) \mu \nonumber\\
&=:\ell_2(\boldsymbol{\nu}, \mathbf{n} ,\bgamma)\,,
\label{SecondLagrangian}
\end{align}
where we allow for $\gamma=-({\nabla}\chi)\,\chi^{-1}\in\Omega^1(\mathcal{D},\mathfrak{so}(3))$ to be an extra dynamical variable, we denote by $\boldsymbol{\gamma}=
( \boldsymbol{\gamma}_1, \boldsymbol{\gamma}_2, \boldsymbol{\gamma}_3)$ the corresponding $\mathbb{R}^3$-valued one-form, $\boldsymbol{\gamma}_i\in \Omega^1( \mathcal{D})$, $i=1,2,3$, and $\bgamma\times \mathbf{n} \in\Omega^1(\mathcal{D},\mathbb{R}^3)$ is defined by
\begin{equation}\label{def_cros}
(\bgamma\times \mathbf{n} )(v_x)=\bgamma(v_x)\times \mathbf{n},\quad v_x\in T_x\mathcal{D},
\end{equation}
or, in local coordinates, $\bgamma\times \mathbf{n}=\left(\bgamma_i\times \mathbf{n}\right)\de x^i$.
In
Appendix \ref{Appendix} it is shown how the Frank energy is written in terms of ${\bf n}$ and $\gamma$.

It is important to notice that the expression for $L=L( \chi , \dot \chi , \mathbf{n} _0, \gamma _0 )$ may not be defined when $ \gamma_0 \neq 0$. In this case  $\ell_2$ is only defined on the orbit of $ \gamma_0 =0$, that is, on $ \gamma $ of the form $ \gamma 
=-({\nabla} \chi)\chi^{-1}$. However, this does not affect the reduction process, as long as the expression $L( \chi , \dot \chi , \mathbf{n} _0,0)$ is invariant under the isotropy group of $\gamma_0=0$. It is interesting to observe that this construction is identical to the reduction process occurring for the dynamics of polymer chains, see \cite{ElGBHoPuRa,GBHoRa2011} to which we also refer for more details about the reduction processes when $\gamma_0=0$. 

\begin{remark}[Symmetry breaking and isotropy subgroup]
{\rm Notice that, in the context of symmetry breaking \cite{GBTr2010}, the above reduction is no longer performed with respect to the isotropy subgroup of $\mathbf{e}_3$, i.e.,  $\mathcal{F(D},S^1)=\mathcal{F}(\mathcal{D},SO(3))_{\mathbf{e}_3\,}$, as it happens for Ericksen-Leslie dynamics. Rather, since the isotropy subgroup of $\gamma_0=0$ is given by $SO(3)\subset\mathcal{F(D},SO(3))$, the entire reduction process is with respect to the isotropy subgroup 
\[
\mathcal{F}(\mathcal{D},SO(3))_{(\mathbf{e}_3,0)}=\mathcal{F(D},S^1)\cap SO(3)=S^1
\,.
\]
See Section \ref{sec:GeneralCase} for a more detailed discussion on this topic. \quad $\blacklozenge$}
\end{remark}

At this point, the resulting associated Euler-Poincar\'e variational principle is
\begin{equation}\label{HamPrinc2}
\delta\int^{t_1}_{t_0}\ell_2(\bnu,\bn,\bgamma)\,\de t=0\,,
\end{equation}
subject to the variations $\delta\bnu=\partial_t{\boldsymbol\eta}+\bnu\times\boldsymbol\eta$ and $\delta(\bn,\bgamma)=\left(\boldsymbol\eta\times\bn,-{\nabla}^{\bgamma\,}\boldsymbol\eta\right)$ for $\boldsymbol{\eta}(t_0)=
\boldsymbol{\eta}(t_1) = 0$. 
Here $\boldsymbol{\hat{\eta}}=(\delta\chi)\chi^{-1}$ and ${\nabla}^\bgamma$ denotes the covariant differentiation 
\[
{\nabla}^{\boldsymbol{\bgamma}}\mathbf{a}:={\nabla}\mathbf{a}+\bgamma\times\mathbf{a} \in\Omega^1(\mathcal{D}, 
\mathbb{R}^3)\,,
\] 
for any $\mathbf{a} \in \mathcal{F}(\mathcal{D}, \mathbb{R}^3)$.
Then, the affine Euler-Poincar\'e equations are
\begin{equation}\label{EL_gamma}
\left\{
\begin{array}{l}
\vspace{0.2cm}\displaystyle\frac{d}{dt}\frac{\delta \ell _2 }{\delta \boldsymbol{\nu}}=\boldsymbol{\nu}\times \frac{\delta \ell _2 }{\delta \boldsymbol{\nu}}+\operatorname{div}^\bgamma\frac{\delta\ell_2}{\delta\bgamma} +\mathbf{n} \times \frac{\delta \ell _2 }{\delta \mathbf{n} }\\
\vspace{0.2cm}\displaystyle \partial_ t \mathbf{n} + \mathbf{n} \times \boldsymbol{\nu }=0\\
\displaystyle \partial _t \bgamma +  \bgamma \times \boldsymbol{\nu}+ {\nabla} \boldsymbol{\nu}=0,\quad{\bgamma}_0=0\,,
\end{array}\right.
\end{equation}
where we have introduced the covariant divergence of
$\boldsymbol{\kappa} \in \mathfrak{X}(\mathcal{D}, 
\mathbb{R}^3)$ by
\[
\operatorname{div}^\bgamma\boldsymbol\kappa:=\operatorname{div}\boldsymbol\kappa+\operatorname{Tr} \!\left( \boldsymbol{\gamma } \times \boldsymbol\kappa  \right) 
=\frac{\partial\boldsymbol\kappa^i}{\partial x^i}+{\boldsymbol\gamma_i}  \times\boldsymbol\kappa^i \in \mathcal{F}(\mathcal{D}, \mathbb{R}^3)
\,.
\]
Note that the above definition follows, as usual, from the
requirement that the operators $\nabla^\bgamma$ and 
$\operatorname{div}^\bgamma$ are related by
\[
\left\langle  \boldsymbol{\kappa}, \nabla^\bgamma \mathbf{a}
\right\rangle : = \int_\mathcal{D} \left(\nabla^\bgamma \mathbf{a}\right)( \boldsymbol{\kappa})\, \mu = 
- \int_\mathcal{D} \mathbf{a} \cdot 
\operatorname{div}^\bgamma\boldsymbol{\kappa}\, \mu  = :
- \left\langle \operatorname{div}^\bgamma\boldsymbol{\kappa}, 
\mathbf{a} \right\rangle.
\]
Notice that, the final form of the Lagrangian \eqref{SecondLagrangian} yields the equations of motion
\begin{equation}\label{EL_Eringen-gamma}
\left\{
\begin{array}{l}
\vspace{0.2cm}\displaystyle J\bn\times(\bn\times\partial_t{\bnu})=
J(\bn\times(\bn\times{\bnu}))\times\bnu+
\frac{\partial}{\partial x^i}
\frac{\partial \Phi}{\partial\bgamma_i}+
{\boldsymbol\gamma_i}  \times
\frac{\partial \Phi}{\partial\bgamma_i}+
\mathbf{n} \times \frac{\partial \Phi}{\partial \mathbf{n} }\\
\vspace{0.2cm}\displaystyle \partial_ t \mathbf{n} + \mathbf{n} \times \boldsymbol{\nu }=0,\\
\displaystyle \partial _t \bgamma +  \bgamma \times \boldsymbol{\nu}+ {\nabla} \boldsymbol{\nu}=0,\quad{\bgamma}_0=0\,,
\end{array}\right.
\end{equation}
where the free energy  $\Phi(\bn,\bgamma)$ is defined in terms of the Frank energy $F(\bn,\nabla\bn)$ in \eqref{standard_energy} as
\begin{equation}\label{Frank-variant}
\Phi(\bn,\bgamma):=F(\bn,\bn\times\bgamma)
\,,
\end{equation}
through the relation $\nabla\bn=\bn\times\bgamma$ (see Appendix \ref{Appendix}).

Upon using \eqref{SecondLagrangian}, combining the first two equations yields
\begin{equation}
J\partial_t^2\mathbf{n} -\left(\mathbf{n\cdot h}+J\,\mathbf{n}\cdot\partial_t^2\mathbf{n}\right)\mathbf{n}+\mathbf{h}=0,
\label{ericksenleslie2}
\end{equation}
where the molecular field is expressed as
\begin{equation}
\mathbf{h}=\frac{\partial \Phi}{\partial \bn}-\bn\times\left(\frac{\partial}{\partial x^i}\frac{\partial \Phi}{\partial \bgamma_i}+ \boldsymbol{\gamma }_i \times \frac{\partial  \Phi }{\partial  \boldsymbol{\gamma}_i }  \right).
\label{molecularfield2}
\end{equation}
Notice that, contrarily to what happens for \eqref{ELeslie}, the dynamics of $\bnu$ must be computed to evaluate  the instantaneous value of $\Phi$ and $\mathbf{h}$.

Thus, if we suppose that the reference director field $ \mathbf{n} _0 $ is constant, then \eqref{EL} and \eqref{EL_gamma} are equivalent since they are induced by the \emph{same} Euler-Lagrange equations for $ \mathcal{L}(\chi,\dot\chi) $ on $T \mathcal{F} ( \mathcal{D} , SO(3))$. We shall verify this fact explicitly in the next
subsection.
\begin{remark}[The initial condition $\bgamma_0$]{\rm 
When one allows for $\bgamma_0\neq0$, the reduced equations in \eqref{EL_gamma} still make sense, thereby extending EL dynamics to account for disclination dynamics. Notice that in this case, equations \eqref{EL_gamma} still preserve the relation $\nabla\bn-\bn\times\bgamma=0$, since 
\[
\left(\frac{\partial}{\partial t}-\bnu\times\right)(\nabla\bn-\bn\times\bgamma)=0\,.
\]
Thus, the initial conditions   $\bgamma_0$ and $\bn_0$ may be strictly related through the relation $\nabla\bn_0=\bn_0\times\bgamma_0$. It is important to emphasize that the projection  $\bn_0\cdot\bgamma_0$ gives zero contribution to the gradient $\nabla\bn_0$. Then, Eringen's expression of the wryness tensor $\bgamma_0=\nabla\bn_0\times\bn_0$ (that is $\bn_0\cdot\bgamma_0=0$) becomes a convenient initial condition, which is not preserved in time. } \quad $\blacklozenge$
\end{remark}

\subsection{Compatibility}

Upon choosing $\bn=\chi\,\bn_0$ and $\boldsymbol{\widehat{\gamma}}=-({\nabla}\chi)\,\chi^{-1}$, the induced variational principles \eqref{HamPrinc1} and \eqref{HamPrinc2} must be the same:
\[
\delta\int^{t_1}_{t_0}\ell_1(\bnu,\bn)\,\de t=\delta\int^{t_1}_{t_0}L_{\mathbf{n}_0}(\chi,\dot{\chi})\,\de t=\delta\int^{t_1}_{t_0}L_{(\mathbf{n}_0,0)}(\chi,\dot{\chi})\,\de t=\delta\int^{t_1}_{t_0}\ell_2(\bnu,\bn,\bgamma)\,\de t
\]
for any variation of $\chi$ vanishing at the endpoints and $\delta(\bn,\bgamma)=\left(\boldsymbol\eta\times\bn,-{\nabla}^{\bgamma\,}\boldsymbol\eta\right)$, where $\boldsymbol{\widehat{\eta}}=(\delta\chi)\chi^{-1}$.
Thus, {\color{black} upon denoting by $\langle\cdot\,,\cdot\rangle$ the pairing between vectors and convectors on either the director space $\mathcal{F(D,}S^2)$ or the space $\Omega ^1( \mathcal{D},\mathfrak{so}(3))$,} one has
\[
\int^{t_1}_{t_0}\left\langle\frac{\delta\ell_1}{\delta\bn},\delta\bn\right\rangle\de t=\int^{t_1}_{t_0}\left(\left\langle\frac{\delta\ell_2}{\delta\bn},\delta\bn\right\rangle+\left\langle\frac{\delta\ell_2}{\delta\bgamma},\delta\bgamma\right\rangle\right)\de t
\]
so that
\[
\int^{t_1}_{t_0}\left\langle\frac{\delta\ell_1}{\delta\bn},\boldsymbol\eta\times\bn\right\rangle\de t=\int^{t_1}_{t_0}\left(\left\langle\frac{\delta\ell_2}{\delta\bn},\boldsymbol\eta\times\bn\right\rangle-\left\langle\frac{\delta\ell_2}{\delta\bgamma},\de^\bgamma \boldsymbol\eta\right\rangle\right)\de t,
\]
where we ignore variations in $\bnu$ since they give equal contributions which cancel each other because $\delta \ell_1/\delta\bnu=\delta \ell_2/\delta\bnu$. In conclusion, isolating $\boldsymbol\eta$ yields
\begin{equation}\label{compatibility}
\bn\times\frac{\delta\ell_1}{\delta\bn}=\bn\times\frac{\delta\ell_2}{\delta\bn}+\operatorname{div}^\bgamma\frac{\delta\ell_2}{\delta\bgamma} 
\end{equation}
which can be used to show that \eqref{EL_gamma} is compatible with \eqref{EL}. Indeed, we check that equations \eqref{EL_gamma} still produce
\begin{equation}\label{conservation}
\frac{\partial}{\partial t}\left(\frac{\delta\ell_2}{\delta\bnu}\cdot\bn\right)=J\,\frac{\partial}{\partial t}\left(\bnu\cdot\bn\right)=0\,,
\end{equation}
since by \eqref{compatibility} all terms involving $\delta \ell_2/\delta\gamma$ are orthogonal to $\bn$ and give zero contribution. Notice that the above constant is actually a momentum map arising from the invariance of EL dynamics under the isotropy group $\mathcal{F}(\mathcal{D},S^1)$ of $\mathbf{e}_3$; see \cite{GBTr2010} and also \eqref{Noether1} below. The conservation of $\bn\cdot\delta\ell_2/\delta\bnu$ is then to be considered a conservation law arising from Noether's theorem, which is inherited from the system \eqref{EL}. {\color{black}In conclusion, we have proven the following result.
\begin{theorem}
\label{first_compatibility_thm}
The Ericksen-Leslie equations \eqref{ELeslie} {\rm (}with the molecular field \eqref{molecularfield1}{\rm )} are equivalent to equations \eqref{EL_Eringen-gamma} {\rm (}with free energy given by \eqref{Frank-variant}{\rm )}.
\end{theorem}
\paragraph{Proof.} As we have seen, systems \eqref{EP_form_EL} 
and \eqref{EL_Eringen-gamma} are the Euler-Poincar\'e equations associated 
to the Lagrangians \eqref{ELlagrangian1} and \eqref{SecondLagrangian}, respectively. In turn, these two Lagrangians arise by
Euler-Poincar\'e reduction relative to two different symmetry
groups from the same Lagrangian \eqref{UnreducedLagrangian} in material representation. Therefore, using \eqref{compatibility} we conclude
\begin{equation}
\label{identity_useful_EL}
\mathbf{n} \times  \mathbf{h}
= \frac{\partial}{\partial x^i}\frac{\partial \Phi}{\partial\bgamma_i} + {\boldsymbol\gamma_i}  \times\frac{\partial \Phi}{\partial\bgamma_i} + \mathbf{n} \times \frac{\partial \Phi}{\partial \mathbf{n} }\,,
\end{equation}
thereby transforming the first equation in 
\eqref{EL_Eringen-gamma} into the first equation in 
\eqref{EP_form_EL}. Thus, the $\boldsymbol{\gamma}$-equation
decouples in \eqref{EL_Eringen-gamma}. Since we already
showed in Section \ref{sec:ErLe} that \eqref{ELeslie} is equivalent to \eqref{EP_form_EL},
the statement follows.
\quad $\blacksquare$
\medskip

Notice that, if $\gamma_0\neq0$, the uniaxial property $\bnu_0\cdot\bn_0=0$ is not preserved, since \eqref{compatibility} is no longer true. This indicates that 
changes occur in the molecular configuration of the system. Thus, the director parameter must be replaced by a suitable inertia tensor, which becomes the new order parameter field. This is precisely what happens in the Landau-deGennes dynamics of the alignment tensor \cite{deGennes1971}. This treatment is the basis of the micropolar theory of liquid crystals, which was pioneered by Eringen \cite{Eringen1993, Eringen1997}. However, before approaching this problem, we shall show how the two constructions presented in this section are actually special cases of two reduction processes that can be carried out for any continuous medium with broken symmetry.

\section{Reductions for continua with broken symmetry}\label{sec:GeneralCase}

This section generalizes the two constructions previously applied to nematic liquid crystals to arbitrary continuum systems with broken symmetry. More precisely, the rotation group $SO(3)$ for the orientational order is replaced by an arbitrary Lie group $\mathcal{O}$ acting transitively on an order parameter manifold $M$ and
$\mathfrak{so}(3)\cong \mathbb{R}^3$ by the Lie algebra 
$\mathfrak{o}$ of $\mathcal{O}$.
Then, if $n_0\in M$ is a given order parameter variable, it follows that $M$ is the coset manifold $\mathcal{O/P}$, where $\mathcal{P}:=\mathcal{O}_{n_0}\subset\mathcal{O}$ is the isotropy subgroup fixing $n_0${\color{black}, i.e.
\[
\mathcal{O}_{n_0}:=\left\{\chi\in\mathcal{O}\ |\ \chi n_0=n_0\right\}
.
\] Here, the concatenation notation $\chi n_0$ is used for the  $\mathcal{O}$-action on $M$. The explicit expression of this group action depends on the special case under consideration.} This is precisely the same setting as in \cite{GBTr2010}.

For continuous media, one replaces $\mathcal{O}$ and $M$ by $\mathcal{F(D,O)}$ and $\mathcal{F(D},M)$, where $\mathcal{D}$ is the spatial domain of the medium, so that $\mathcal{F(D,O})_{n_0}=\mathcal{F(D,P)}\subset\mathcal{F(D,O)}$, where $n_0\in M$ is identified with a constant function on $\mathcal{D}$. Note that here we consider the action of $\mathcal{F(D,O)}$ on $\mathcal{F(D},M)$ naturally induced
by the $\mathcal{O}$-action on $M$.
At the fundamental unreduced level, one starts with a Lagrangian functional $\mathcal{L}:T\mathcal{F(D,O)}\to\Bbb{R}$, which is typically of the type
\begin{equation}
\label{primitive_lagrangian}
\mathcal{L}(\chi,\dot\chi)=\int_\mathcal{D}\mathscr{L}(\chi,\dot\chi,\chi n_0,{\nabla}(\chi n_0))\,\mu
\end{equation}
so that the fixed order parameter $n_0$ appears in the Lagrangian density $\mathscr{L}$ through both $\chi n_0$ and its differential ${\nabla}(\chi n_0)$. In all cases under consideration, the Lagrangian $\mathcal{L}$ possesses the following invariance properties:
\begin{align}\label{invariance1}
\mathcal{L}(\chi,\dot\chi)=
\int_\mathcal{D}\mathscr{L}(\chi,\dot\chi,\chi n_0,{\nabla}(\chi n_0))\,\mu&=
\int_\mathcal{D}\mathscr{L}(\dot\chi\chi^{-1},\chi n_0,{\nabla}(\chi  n_0))\,\mu=\ell_1(\nu,n)
\end{align}
and, if $n_0$ is constant in space,
\begin{align}\label{invariance2}
\mathcal{L}(\chi,\dot\chi)&=
\int_\mathcal{D}\mathscr{L}(\chi,\dot\chi,\chi n_0,{\nabla}(\chi n_0))\,\mu =
\int_\mathcal{D}\mathscr{L}(\dot\chi\chi^{-1},\chi n_0,(({\nabla}\chi)\chi^{-1})_M (\chi n_0))\,\mu \nonumber \\
&=:\ell_2(\nu,n,\gamma)\,,
\end{align}
for $\nu:=\dot{\chi}\chi^{-1}$, $n:=\chi n_0$ and 
$\gamma:=-({\nabla}\chi)\chi^{-1}$. Here $(({\nabla}\chi)\chi^{-1})_M (\chi n_0):T\mathcal{D}\rightarrow TM$ is defined by
\[
(({\nabla}\chi)\chi^{-1})_M (\chi n_0)(v_x)=\left({\nabla}\chi(v_x)\chi(x)^{-1}\right)_M(\chi(x)n_0(x))\in T_{n(x)}M,\;\; v_x\in T_x\mathcal{D},
\]
where $\xi_M\in\mathfrak{X}(M)$ denotes the infinitesimal generator associated to the Lie algebra element $\xi\in\mathfrak{o}$. Note that in the formula above we have $\xi={\nabla}\chi(v_x)\chi(x)^{-1}\in\mathfrak{o}$. 
Each of these invariance properties involves a distinct reduction procedure that, in turn, produces different Euler-Poincar\'e equations of motion. As is explained below, these two reduced systems are compatible since they arise from the \emph{same} unreduced Lagrangian $\mathcal{L}(\chi,\dot{\chi})$.

\subsection{First reduction} 
\label{sec_first_reduction}
This reduction procedure is based on the invariance property \eqref{invariance1} and it follows  precisely the same steps as in Section 2.1 of \cite{GBTr2010} (see theorem 2.1 therein). In particular, this reduction is performed with respect to the isotropy subgroup $\mathcal{F(D,O)}_{n_0}=\mathcal{F}\!\left(\mathcal{D,O}_{\!n_0}\right)$, because of the diffeomorphism 
\begin{align*}
T\mathcal{F(D,O)}/\mathcal{F}\!\left(\mathcal{D,O}_{\!n_0}\right)&\longrightarrow\mathcal{F(D},\mathfrak{o})\times\mathcal{F(D},M)
\\
\left[(\chi,\dot\chi)\right]&\, \mapsto\left(\dot{\chi}\chi^{-1\,},\chi n_0\right)\,;
\end{align*}
see \cite[Remark 2.5 and equation (3.1)]
{GBTr2010}.

Since the invariance property \eqref{invariance1}
implies
\[
\delta\int_{t_0}^{t_1}\mathcal{L}(\chi,\dot{\chi})\,\de t=\delta\int_{t_0}^{t_1}\ell_1(\nu,n)\,\de t=0\,,
\]
the Euler-Poincar\'e variational principle for $\ell_1$ involves the variations
\begin{align*}
\delta\nu=&\delta\!\left(\dot{\chi}\chi^{-1}\right)=\partial_t{\eta}+\left[\nu,\eta\right]
\\
\delta n=&\delta\!\left(\chi n_0\right)=\eta_M (n),
\end{align*}
where $\eta:=(\delta\chi)\chi^{-1} \in \mathcal{F}( \mathcal{D}, \mathfrak{o})$ and the dot notation stands for partial time derivative. Here the index $M$ on 
$\eta_M( n)$ for $\eta\in\mathcal{F}(\mathcal{D},
\mathfrak{o})$ and $n\in\mathcal{F}(\mathcal{D},M)$, denotes the infinitesimal generator of the 
$\mathcal{F(D,O)}$-action on $\mathcal{F(D},M)$, which is formally given at $x\in\mathcal{D}$ by 
$\eta(x)_{M}(n(x))$. Thus, the resulting equations of motion read
\begin{equation}\label{EP_continuum}
\frac{\partial}{\partial t}\frac{\delta
\ell_1}{\delta\nu}+\operatorname{ad}^*_\nu\frac{\delta \ell_1}{\delta\nu}=\mathbf{J}\!\left(\frac{\delta
\ell_1}{\delta n}\right),\quad \partial_t n=\nu_M( n)\,,
\end{equation}
where $\mathbf{J}:T^*\mathcal{F(D},M)\rightarrow\mathcal{F(D},\mathfrak{o}^*)$ is the momentum map of the cotangent lifted action of $\mathcal{F(D,O)}$ on $\mathcal{F(D},M)$, which is given by 
\cite[Theorem 12.1.4]{MaRa1999}
\[
\left\langle\mathbf{J}(\alpha_n),\xi\right\rangle=\left\langle\alpha_n,\xi_M(n)\right\rangle\,\qquad\forall \,\alpha_n\in T^*_n \mathcal{F(D},M)
\,,
\]
and $\langle\cdot\,,\cdot\rangle$ denotes the pairing between vectors and covectors on either the gauge Lie algebra $\mathcal{F(D},\mathfrak{o})$ or the order parameter space $\mathcal{F(D},M)$.

Notice that $\mathcal{F}\!\left(\mathcal{D,O}_{\!n_0}\right)$-invariance of $\mathcal{L}(\chi,\dot{\chi})=\ell_1(\nu,n)$ yields, by Noether's theorem, the following momentum map conservation (see \cite[\S4]{GBTr2010})
\begin{equation}\label{Noether1}
\frac{\partial}{\partial t}\!\left(i^*\!\left(\operatorname{Ad}^*_{\chi}\frac{\delta \ell_1}{\delta \nu}\right)\right)=0\,,
\end{equation}
where $i^*$ is the dual of the Lie algebra inclusion $i:\mathcal{F(D},\mathfrak{o}_{n_0})\to \mathcal{F(D},\mathfrak{o})$. This is the immediate generalization of the relation \eqref{conservation} for uniaxial nematics, for which $\mathcal{O}_{n_0}=S^1$ and $i(r)=(0,0,r)$. The above conserved quantity is readily seen to arise as a momentum map $\mathcal{J}:T^*\mathcal{F(D,O)}\to\mathcal{F} \left(\mathcal{D},\mathfrak{o}_{n_0}^*\right)$ by the following computation
\[
\left\langle\mathcal{J}(\alpha_\chi),\zeta\right\rangle=\left\langle\alpha_\chi,\zeta_\mathcal{O}(\chi)\right\rangle=\left\langle\alpha_\chi,\chi i(\zeta)\right\rangle=\left\langle\chi^{-1}\alpha_\chi, i(\zeta)\right\rangle=\left\langle i^*\!\left(\chi^{-1}\alpha_\chi\right),\zeta\right\rangle
\]
where $\chi^{-1}\alpha_\chi=\operatorname{Ad}^*_{\chi}(\alpha_{\chi\,}\chi^{-1})$ and $\zeta\in\mathcal{F(D},\mathfrak{o}_{n_0})$ is arbitrary. The index $\mathcal{O}$ on $\zeta_\mathcal{O}$ denotes the infinitesimal generator of the right $\mathcal{F}\!\left(\mathcal{D,O}_{\!n_0}\right)$-action on $\mathcal{F(D,O)}$. This Lie algebra action is given by ${\chi\mapsto\chi\,i(\zeta)}$.

\subsection{Second reduction}
\label{sec_second_reduction}

In this section we restrict all considerations to a given initial condition $n_0\in M\subset\mathcal{F(D},M)$ (i.e., $n_0$ spatially constant) in order to perform the reduction arising from the invariance property \eqref{invariance2}. The construction in this section is based on the treatment in \cite{GBRa2009}, involving affine actions of the gauge group $\mathcal{F(D,O)}$.

Property \eqref{invariance2} arises mainly from the following observation:
\begin{equation}\label{GenFormulaGradient}
{\nabla} n={\nabla} (\chi n_0)= ({\nabla} \chi) n_0= \left(({\nabla} \chi) \chi^{-1}\right)_{\!M}\chi n_0=\left(({\nabla} \chi) \chi^{-1}\right)_{\!M}n=:-\gamma_M (n)\,,
\end{equation}
which defines the connection one form 
$\gamma:=-({\nabla}\chi)\chi^{-1}\in\Omega^1(\mathcal{D},\mathfrak{o})$. Here, $\gamma_M(n)(v_x): = \left(\gamma(v_x)\right)_M(n(x)) \in T_{n(x)}M$ for any 
$v_x \in T_x\mathcal{D}$. Thus, it becomes natural to incorporate $\gamma$ in the equations of motion as an extra dynamical variable. This step requires precisely the reduction given by the invariance 
\eqref{invariance2}, and hence we conclude
\begin{align}
\mathcal{L}(\chi,\dot\chi)&=
\int_\mathcal{D}\mathscr{L}(\dot\chi\chi^{-1},\chi n_0,(({\nabla}\chi)\,\chi^{-1})_M (\chi n_0))\,\mu
=
\int_\mathcal{D}\mathscr{L}(\nu,n,-\gamma_M( n))\,\mu
\nonumber \\
&=:\ell_2(\nu,n,\gamma).
\label{invariance2-bis}
\end{align}
In this case, the reduction proceeds with respect to the isotropy subgroup of $(n_0(x),\gamma_0(x))=(n_0,0)$, which is necessarily a subgroup of $\mathcal{F(D,O)}_{n_0}$. More precisely, since $\mathcal{F(D,O)}_{n_0}=\mathcal{F}\!\left(\mathcal{D,O}_{\!n_0}\right)$ and $\mathcal{F(D,O)}_{\gamma_0=0}=\mathcal{O}$ one has 
\[
\mathcal{F(D,O)}_{(n_0,0)}=\mathcal{F}\!\left(\mathcal{D,O}_{\!n_0}\right)\cap \mathcal{O}=\mathcal{O}_{\!n_0}\,.
\]
Thus, the second invariance property \eqref{invariance2} leads to a reduction involving the isotropy group $\mathcal{O}_{\!n_0}$, which is much smaller than the isotropy $\mathcal{F}\!\left(\mathcal{D,O}_{n_0}\right)$ used in the first reduction presented in 
\S\ref{sec_first_reduction} arising from the invariance
property  \eqref{invariance1}.

Notice that, upon considering the gauge action of $\mathcal{F(D,O)}$ on $\Omega^1(\mathcal{D},\mathfrak{o})$
\begin{equation}\label{affineaction}
 \gamma _0  \mapsto  \psi ^{-1} \gamma _0 \psi + \psi ^{-1} {\nabla} \psi\,,
\end{equation}
the invariance property \eqref{invariance2} takes the form
\begin{align*}
\mathcal{L}(\chi,\dot\chi)=
\int_\mathcal{D}\mathscr{L}\big(\dot\chi\chi^{-1},\chi n_0,-(\chi^{-1}\gamma_0)_M (\chi n_0)\big)\,\mu
&=
\int_\mathcal{D}\mathscr{L}(\nu,n,-\gamma_M (n))\,\mu
=:\ell_2(\nu,n,\gamma)
\end{align*}
where $\gamma_0=0$ is a fixed initial condition.
At this point, since the invariance property \eqref{invariance2}
implies
\[
\delta\int_{t_0}^{t_1}\mathcal{L}(\chi,\dot{\chi})\,\de t=\delta\int_{t_0}^{t_1}\ell_2(\nu,n,\gamma)\,\de t=0\,,
\]
the Euler-Poincar\'e variational principle 
for $\ell_2$ 
involves the variations
\begin{align*}
\delta\nu=&\delta\!\left(\dot{\chi}\chi^{-1}\right)=\partial_t{\eta}+\left[\nu,\eta\right]
\\
\delta n=&\delta\!\left(\chi n_0\right)=\eta_M (n)
\\
\delta\gamma=&\delta\!\left(\chi^{-1} \gamma_0\right)=-\eta_{\Omega^1}(\gamma_0)=-{\nabla}^\gamma\eta,
\end{align*}
where ${\nabla}^\gamma\lambda:={\nabla}\lambda+\left[\gamma,\lambda\right] \in\Omega^1(\mathcal{D}, \mathfrak{o})$ is the covariant differential of $\lambda \in \mathcal{F}(\mathcal{D}, \mathfrak{o})$ and the subscript 
${\Omega^1}$ on $\eta_{\Omega^1}$ denotes the infinitesimal generator of the affine action \eqref{affineaction}. In the variations above,
$\eta$ is a path in $\mathcal{F}(\mathcal{D}, \mathfrak{o})$  vanishing at $t_0$ and $t_1$.

The resulting equations of motion are
\begin{equation}
\label{EP_continuum2}
\begin{aligned}
&\frac{\partial}{\partial t}\frac{\delta
\ell_2}{\delta\nu}+\operatorname{ad}^*_\nu\frac{\delta \ell_2}{\delta\nu}=\mathbf{J}\!\left(\frac{\delta
\ell_2}{\delta n}\right)+\operatorname{div}^{\gamma\!}\left(\frac{\delta \ell_2}{\delta \gamma}\right),\\
&(\partial_t n,\partial_t\gamma)=\left(\nu_M (n),-{\nabla}^\gamma\nu\right), \quad \gamma_0=0.
\end{aligned}
\end{equation}
The covariant divergence is now written as
\[
\operatorname{div}^\gamma\left(\frac{\delta \ell_2}{\delta \gamma}\right)
:=\operatorname{div}\frac{\delta\ell_2}{\delta\gamma}-\operatorname{Tr} \left( \mathrm{ad}^*_{\gamma}\frac{\delta\ell_2}{\delta\gamma} \right) = \frac{\partial}{\partial x^i}
\left(\frac{\delta \ell_2}{\delta \gamma_i}\right) + 
\mathrm{ad}^*_{\gamma_i}\frac{\delta\ell_2}{\delta\gamma_i}
\in \mathcal{F}(\mathcal{D}, \mathfrak{o}^\ast);
\]
this follows from the defining relation
\[
\left\langle  \kappa, \nabla^\bgamma\zeta
\right\rangle : = \int_\mathcal{D} \left(\nabla^\bgamma \zeta\right)(\kappa)\, \mu = 
- \int_\mathcal{D} \left\langle  
\operatorname{div}^\bgamma\kappa, \zeta\right\rangle\, \mu  = :
- \left\langle \operatorname{div}^\bgamma\kappa, 
\zeta \right\rangle,
\]
for any $\zeta \in \mathcal{F}(\mathcal{D}, \mathfrak{o})$
and $\kappa \in \mathfrak{X}(\mathcal{D}, \mathfrak{o}^\ast)$, 
where the pairing in the second integrand is the duality
pairing $\left\langle\,, \right\rangle: \mathfrak{o}^\ast
\times \mathfrak{o} \rightarrow\mathbb{R}$. 
Here, $\mathbf{J}:T^*\mathcal{F(D},M)\rightarrow\mathcal{F(D},\mathfrak{o}^*)$ is the same momentum map as in
\S\ref{sec_first_reduction}, while $\mathbf{K}(\gamma,w)
:=\operatorname{div}^\gamma w$ is the momentum map 
$\mathbf{K}:T^*\Omega^1(\mathcal{D},\mathfrak{o})\rightarrow
\mathcal{F(D},\mathfrak{o}^*)$ induced by the cotangent lifted 
action of $\mathcal{F(D,O)}$ on 
$\Omega^1(\mathcal{D},\mathfrak{o})$.

\subsection{Compatibility of the two approaches}
Since the two approaches arise from the \emph{same} unreduced Lagrangian, they are compatible. This compatibility is reflected in the following relations
\[
\frac{\partial}{\partial t}\frac{\delta
\ell_1}{\delta\nu}+\operatorname{ad}^*_\nu\frac{\delta \ell_1}{\delta\nu}-\mathbf{J}\!\left(\frac{\delta
\ell_1}{\delta n}\right)=\frac{\partial}{\partial t}\frac{\delta
\ell_2}{\delta\nu}+\operatorname{ad}^*_\nu\frac{\delta \ell_2}{\delta\nu}-\mathbf{J}\!\left(\frac{\delta
\ell_2}{\delta n}\right)-\operatorname{div}^{\gamma\!}\left(\frac{\delta \ell_2}{\delta \gamma}\right)=0
\]
which arise from the variational principles
\[
\delta\int^{t_1}_{t_0}\ell_1(\nu,n)\,\de t=\delta\int^{t_1}_{t_0}\ell_2(\nu,n,\gamma)\,\de t=0.
\]
In the particular case when $\delta \ell_1/\delta\nu=\delta\ell_2/\delta\nu$ we obtain
\begin{equation}
\label{momap_identity}
\mathbf{J}\!\left(\frac{\delta
\ell_1}{\delta n}\right)=\mathbf{J}\!\left(\frac{\delta
\ell_2}{\delta n}\right)+\operatorname{div}^\gamma\left(\frac{\delta \ell_2}{\delta \gamma}\right),
\end{equation}
which generalizes the analogous relation \eqref{compatibility} previously found for nematodynamics. Therefore,
since by construction, the systems \eqref{EP_continuum2} and 
\eqref{EP_continuum} arise from the same unreduced Lagrangian \eqref{primitive_lagrangian}, we obtain the following result:
\begin{theorem}  
Upon using the relation $\nabla n=-\widehat{\gamma}_M(n)$, the equations \eqref{EP_continuum2} and \eqref{EP_continuum} are equivalent.
\end{theorem}
We want to emphasize that many of the subsequent results
are corollaries of this theorem for special Lagrangians.

\subsection{More general Lagrangians}

So far, we considered the case in which the parameter $n_0$ appears in the Lagrangian density $\mathscr{L}$ only through the term $\chi n_0$ and its gradient ${\nabla}(\chi n_0)$. Then we showed how such a Lagrangian possesses the two invariance properties \eqref{invariance1} and \eqref{invariance2}. However, one can consider the more general case of an invariant Lagrangian of the type
\[
\mathcal{L}(\chi,\dot\chi)=
\int_\mathcal{D}\mathscr{L}(\chi,\dot\chi,\chi n_0,{\nabla}\chi )\,\mu=\int_\mathcal{D}\mathscr{L}(\dot\chi\chi^{-1},\chi n_0,({\nabla}\chi)\chi^{-1} )\,\mu=\ell_2(\nu,n,\gamma)\,,
\]
which has a free dependence on the variable $\gamma$, still possessing the initial condition $\gamma_0=0$. In this case, the only invariance property is of the type \eqref{invariance2} and there is no reduction other than that of second type. As above, one regards $\mathcal{L}$ as a Lagrangian $L(\chi,\dot\chi,n_0,0)$ invariant under the isotropy group of $\gamma_0=0$. A simple concrete example of such a situation is when $n_0=(0,0,0)\in\Bbb{R}^3$ and $\mathcal{O}=SO(3)$, which produces the framework for spin glass dynamics \cite{Volovick1980,HoKu1988}. Then, the momentum map associated to the residual $SO(3)$-symmetry (recall that $SO(3)\subset\mathcal{F(D},SO(3))$) is $\mathcal{J}:T^*\mathcal{F}(\mathcal{D},SO(3))\rightarrow\mathfrak{so}(3)^*$, $\mathcal{J}(\alpha_\chi)=\int_\mathcal{D}\chi^{-1}\alpha_\chi\mu$, and therefore yields (by Noether's theorem) the conservation law
\begin{equation}\label{Noether2}
\frac{\de}{\de t}\int_\mathcal{D\!}\left(\operatorname{Ad}^*_{\chi}\frac{\delta\ell_2}{\delta\nu}\right)\mu=\int_\mathcal{D}\operatorname{Ad}^*_{\chi}\!\left(\mathrm{div}^\gamma\frac{\delta\ell_2}{\delta\gamma}\right)\mu=0\,,
\end{equation}
where the second equality follows from the general formula \cite[formula (9.3.7)]{MaRa1999}
\[
\frac{\partial}{\partial t}\left(\mathrm{Ad}^*_\chi\sigma\right)=\mathrm{Ad}^*_\chi\!\left(\dot\sigma+\mathrm{ad}^*_{\dot{\chi}\chi^{-1}}\sigma\right)
\]
and from equations \eqref{EP_continuum2} with $n_0=(0,0,0)=n$. As we have seen, in the case of liquid crystals, the two reductions are both possible (producing $\ell_1$ and $\ell_2$). However, when $\ell_1$ does not exist (e.g., for spin glasses), the only possible conservation law is
\begin{equation}\label{Noether2-bis}
\frac{\de}{\de t}\,j^*\!\left(\int_\mathcal{D}\!\left(\operatorname{Ad}^*_{\chi}\frac{\delta \ell_2}{\delta \nu}\right)\mu\right)=0\,,
\end{equation}
where $j^*$ is the dual of the Lie algebra inclusion $j:\mathfrak{o}_{n_0}\hookrightarrow \mathfrak{o}$. In the case of spin glasses,   $n=(0,0,0)$ yields $\mathfrak{o}_{n_0}=\mathfrak{so}(3)=\mathfrak{o}$, so that $j^*$ reduces to the identity. On the other hand, for liquid crystals $\mathcal{O}_{n_0}=S^1\subset SO(3)=\mathcal{O}$, so that the above conservation law is immediately implied by applying Noether's theorem to $\ell_1$, as we did already in \eqref{Noether1}.

The next section will apply this general setting to the case of microfluids. In this context the order parameter field is the molecule inertia tensor (\emph{microinertia}) taking values in $M=\operatorname{Sym}(3)$, the space of $3\times 3$ symmetric matrices. Both reductions above apply naturally in this context.

\bigskip

\subsection{Reductions for micropolar media}\label{Sec:micropolar}
Micropolar media are continuum media in which the shape of each rigid particle may change in time, depending on the point in space. The molecule shape is given by an appropriate microinertia tensor, which also appears in the expression of the free energy, denoted by $\Phi$. Then, the unreduced Lagrangian is given by
\begin{align}
\label{micropolar_material_lagrangian}
\mathcal{L} ( \chi , \dot \chi )=&\int_\mathcal{D}\mathscr{L}\big(\chi,\dot{\chi},\chi j_0\chi^{-1},{\nabla}(\chi j_0\chi^{-1})\big)\mu
\nonumber \\
=&\frac{1}{2}\int_{ \mathcal{D} } \operatorname{Tr}\left( (i_0 \chi ^{-1} \dot \chi ) ^T \chi ^{-1} \dot \chi  \right)  \mu -\int_{ \mathcal{D} }\Upsilon_1( \chi j_0\chi^{-1},{\nabla} ( \chi j_0\chi^{-1}) ) \mu ,
\end{align}
where $\Upsilon_1$ denotes the free energy, $ j _0 $ is the microinertia  tensor and $i_0:=\frac{1}{2}\operatorname{Tr}(j_0)I_3-j_0$ (or, equivalently, $j_0={i_0-\operatorname{Tr}(i_0)I_3}$). 
Upon repeating exactly the main steps as in the previous sections, one considers $j_0\in\mathcal{F(D},\operatorname{Sym}(3))$ as the order parameter field and, by defining $j=\chi\,j_0\,\chi^{-1}$, one obtains the first reduced Lagrangian
\begin{align}\label{lagrangia}
\ell_1( \boldsymbol{\nu}, j):=&\int_\mathcal{D}\mathscr{L}\big(\dot{\chi}\chi^{-1},\chi j_0\chi^{-1},{\nabla}(\chi j_0\chi^{-1})\big)\mu
=\frac{1}{2} \int_{ \mathcal{D} } (j  \boldsymbol{\nu}) \cdot  \boldsymbol{\nu} \mu -\int_{ \mathcal{D} }\Upsilon_1( j , \nabla j ) \mu,
\end{align}
where 
\[
j_0\mapsto\chi\,j_0\,\chi^{-1}
\]
defines the action of $\mathcal{F(D,}SO(3))$ on $\mathcal{F(D},\operatorname{Sym}(3))$. 

On the other hand, if $j_0$ is constant in space, then we obtain the second reduced Lagrangian
\begin{align}\label{lagrangia2}
\ell_2( \boldsymbol{\nu}, j,\gamma)
:=&\int_\mathcal{D}\mathscr{L}\big(\dot{\chi}\chi^{-1},\chi j_0\chi^{-1},[({\nabla}\chi)\chi^{-1},\chi j_0\chi^{-1}]\big)
\mu \nonumber
\\
=&\ \frac{1}{2} \int_{ \mathcal{D} } (j  \boldsymbol{\nu}) \cdot  \boldsymbol{\nu} \mu -\int_{ \mathcal{D} }
\Upsilon_2(j,\bgamma) \mu,
\end{align}
where $\gamma:=-({\nabla}\chi)\chi^{-1}$ (with $\gamma_0=0$), 
the Lie bracket $\left[\cdot,\cdot\right]$ is the ordinary 
matrix commutator, and $\Upsilon_2(j ,\bgamma):=\Upsilon_1( j ,- [\widehat{\bgamma},j])$. Then, each of the above reduced Lagrangians produces the following equivalent sets of equations, respectively:
\begin{equation}\label{EP3}
\left\{
\begin{array}{l}
\vspace{0.2cm}\displaystyle\frac{\partial}{\partial t}\frac{\delta \ell _1 }{\delta \boldsymbol{\nu}}=\boldsymbol{\nu}\times \frac{\delta \ell _1}{\delta \boldsymbol{\nu}}-\!\overrightarrow{\,\left[
\frac{\delta \ell _1}{\delta j}, j\right]\,} \\
\vspace{0.2cm}\displaystyle\partial_t j+[j,\hat{\boldsymbol{\nu}}]=0
\end{array}\right.
\end{equation}
\begin{equation}\label{EP3-gamma}
\left\{
\begin{array}{l}
\vspace{0.2cm}\displaystyle\frac{\partial}{\partial t}\frac{\delta \ell _2 }{\delta \boldsymbol{\nu}}=\boldsymbol{\nu}\times \frac{\delta \ell _2}{\delta \boldsymbol{\nu}}-\!\overrightarrow{\,\left[\frac{\delta \ell _2}{\delta j}, j\right]\,}+\operatorname{div}^\bgamma\frac{\delta\ell_2}{\delta\bgamma}\\
\vspace{0.2cm}\displaystyle\partial_t j+[j,\hat{\boldsymbol{\nu}}]=0\\
\displaystyle \partial _t \gamma + [ \gamma ,\hat{ \boldsymbol{\nu}}]+ {\nabla} \hat{ \boldsymbol{\nu}}=0\,,\qquad\bgamma_0=0,
\end{array}\right.
\end{equation}
with the notation $\overrightarrow{A\,}_{\!i}:=
-\epsilon_{ijk}A_{jk}$, $A \in \mathfrak{so}(3)$,  for the dual of the hat map \eqref{hatmap}, i.e., for any
$A \in \mathfrak{so}(3)$ and any $\mathbf{a} \in 
\mathbb{R}^3$, we have $\operatorname{Tr}\left(A^T\widehat{\mathbf{a}} \right) = -\operatorname{Tr}\left(A\widehat{\mathbf{a}} \right) = \overrightarrow{A} \cdot \mathbf{a}$.
Upon using the Lagrangians \eqref{lagrangia} and \eqref{lagrangia2}, the  systems \eqref{EP3} and \eqref{EP3-gamma} become
\begin{equation}\label{EP3-A}
\left\{
\begin{array}{l}
\vspace{0.2cm}\displaystyle j\partial_t{\bnu}=
j\bnu\times\bnu+\!\overrightarrow{\,\left[
\frac{\partial \Upsilon_1}{\partial j}, j\right]}-
\!\overrightarrow{\,\left[
\frac{\partial}{\partial x^i}\frac{\partial  \Upsilon_1}{\partial  (\partial_{x^i}j)}, j\right]}\\
\vspace{0.2cm}\displaystyle\partial_t j+[j,\hat{\boldsymbol{\nu}}]=0\,,
\end{array}\right.
\end{equation}
and
\begin{equation}\label{EP3-B}
\left\{
\begin{array}{l}
\vspace{0.2cm}\displaystyle j\partial_t{\bnu}=
j\bnu\times\bnu
-\frac{\partial}{\partial x^i}\frac{\partial \Upsilon_2}{\partial\bgamma_i}+ \boldsymbol{\gamma}^a \times 
\frac{\partial \Upsilon_2}{\partial\boldsymbol{\gamma}^a}\\
\vspace{0.2cm}\displaystyle \partial_t j+[j,\hat{\boldsymbol{\nu}}]=0,\\
\displaystyle \partial _t \bgamma +  \bgamma \times \boldsymbol{\nu}+ {\nabla} \boldsymbol{\nu}=0,\quad{\bgamma}_0=0\,,
\end{array}\right.
\end{equation}
where we have used the identities 
$\widehat{\overrightarrow{A}} = 2A$ for any $A \in \mathfrak{so}(3)$ and
\begin{equation}
\label{objectivity}
\overrightarrow{\left[j,\frac{\partial \Upsilon_2}{\partial j} 
\right]} = \boldsymbol{\gamma}^a \times 
\frac{\partial \Upsilon_2}{\partial\boldsymbol{\gamma}^a} + 
\boldsymbol{\gamma}_i \times 
\frac{\partial \Upsilon_2}{\partial\boldsymbol{\gamma}_i}\,;
\end{equation}
for the notations $\boldsymbol{\gamma}^a \in \Omega^1(\mathcal{D})$ and 
$\boldsymbol{\gamma}_i \in \mathcal{F}( \mathcal{D},\mathbb{R}^3)$ see \eqref{tensor_gamma} and for the
identity above see \cite[Lemma 8.5]{GBRa2009} which is a consequence
of the axiom of objectivity for the free energy $\Upsilon_2$.

\begin{theorem}
\label{equivalence_thm}
Systems \eqref{EP3-A} and \eqref{EP3-B} are equivalent.
\end{theorem}

\paragraph{Proof.} As we have seen, systems \eqref{EP3-A} and 
\eqref{EP3-B} are the Euler-Poincar\'e equations associated 
to the Lagrangians \eqref{lagrangia} and \eqref{lagrangia2}, respectively. In turn, these two Lagrangians arise by
Euler-Poincar\'e reduction relative to two different symmetry
groups from the same Lagrangian \eqref{micropolar_material_lagrangian} in material representation. Therefore, upon replacing $n \in M$ by
$j \in \operatorname{Sym}(3)$ and 
\[
\mathbf{J}\left(\frac{\delta\ell_i}{\delta j} \right) = 
\overrightarrow{\left[j, \frac{\delta\ell_i}{\delta j}\right]} \qquad , i=1,2,
\]
in formula \eqref{momap_identity}, we conclude
\[
\overrightarrow{\,\left[
\frac{\partial \Upsilon_1}{\partial j}, j\right]}-
\!\overrightarrow{\,\left[
\frac{\partial}{\partial x^i}\frac{\partial  \Upsilon_1}{\partial  (\partial_{x^i}j)}, j\right]} = 
-\frac{\partial}{\partial x^i}\frac{\partial \Upsilon_2}{\partial\bgamma_i}+ \boldsymbol{\gamma}^a \times 
\frac{\partial \Upsilon_2}{\partial\boldsymbol{\gamma}^a}\,,
\]
where we have used formula \eqref{objectivity} in the right hand side. Therefore the $\boldsymbol{\gamma}$-equation
decouples in \eqref{EP3-B}, which proves the statement.
\quad $\blacksquare$

\medskip
While \eqref{EP3-A} are the well known equations for micropolar media \cite{Eringen1993, Eringen1997}, the second set of equations \eqref{EP3-B} provide an alternative formulation of the same dynamics, as long as $\bgamma_0=0$. When the latter initial condition is dropped, the two systems are not equivalent and  \eqref{EP3-gamma} yield Eringen's formulation of micropolar liquid crystals, which accounts for disclination dynamics through the disclination density $\boldsymbol{B}$ in \eqref{Magnetic-Field}. Notice that the latter quantity possesses a coordinate-free definition in terms of the Yang-Mills curvature two-form
\[
\boldsymbol{B}(u_x,v_x):=\mathbf{d}\boldsymbol{\gamma}(u_x,v_x)+\bgamma(u_x)\times\bgamma(v_x)\,\qquad u_x,v_x\in T_x\mathcal{D}\,,
\]
where $\mathbf{d}$ denotes the exterior differential. Then, if $\gamma_0\neq0$, the unreduced Lagrangian is
\begin{equation}\label{Eringen_Lagr}
\mathcal{L} ( \chi , \dot \chi )=\frac{1}{2}\int_{ \mathcal{D} } \operatorname{Tr}\left( (i_0 \chi ^{-1} \dot \chi ) ^T \chi ^{-1} \dot \chi  \right)  \mu -\int_{ \mathcal{D} }\Psi( \chi j_0\chi^{-1}, \chi {\nabla}  \chi^{-1}+\chi\gamma_0\chi^{-1}) \mu,
\end{equation}
with some free energy $\Psi$ and the following reduced expression
\begin{equation}\label{reduced_Eringen_Lagr}
\ell_2( \boldsymbol{\nu}, j,\gamma)= \frac{1}{2} \int_{ \mathcal{D} } (j  \boldsymbol{\nu}) \cdot  \boldsymbol{\nu} \mu -\int_{ \mathcal{D} }\Psi( j ,\bgamma) \mu.
\end{equation}
The discussion in the next section shows how this micropolar formulation recovers Ericksen-Leslie nematodynamics if the free energy $\Psi$ equals the Frank  energy \eqref{standard_energy}.

\section{Comparing Eringen and Ericksen-Leslie theories}

\rem{ 
In this section one allows for $\gamma_0\neq0$.

Consider the Lagrangian $\mathcal{L} :T \mathcal{F} ( \mathcal{D} , SO(3)) \rightarrow \mathbb{R}  $ given by
\[
\mathcal{L} ( \chi , \dot \chi )=\frac{1}{2}\int_{ \mathcal{D} } \operatorname{Tr}\left( (i_0 \chi ^{-1} \dot \chi ) ^T \chi ^{-1} \dot \chi  \right)  \mu -\int_{ \mathcal{D} }\Psi( \chi j_0\chi^{-1}, \chi {\nabla}  \chi^{-1}+\chi\gamma_0\chi^{-1}) \mu ,
\]
where $ j _0 $ is the microinertia  tensor and $ \gamma  _0$  is a potential encoding the initial disclination  configuration and $i_0:=\frac{1}{2}\operatorname{Tr}(j_0)I_3-j_0$.

One observes that if $j\mapsto \chi j\chi^{-1}$ then $i\mapsto \chi i\chi^{-1}$.

One easily checks that $ \mathcal{L} $ is invariant under the action of the isotropy subgroup of $( j _0 , \gamma_0 )$. The reduced Lagrangian is
\[
\ell( \boldsymbol{\nu}, j, \gamma )= \frac{1}{2} \int_{ \mathcal{D} } (j  \boldsymbol{\nu}) \cdot  \boldsymbol{\nu} \mu -\int_{ \mathcal{D} }\Psi( j , \gamma ) \mu,
\]
where we used the formula
\[
\operatorname{Tr}((i\nu)^T\nu)=j\boldsymbol{\nu}\!\cdot\!\boldsymbol{\nu},
\]
for  $\nu\in\mathfrak{so}(3)$ and where $j=\operatorname{Tr}(i)\mathbf{I}-i$.
}    

In the previous section, Eringen's micropolar theory was shown to account for disclination dynamics when $\gamma_0\neq0$. Now we shall
show how Eringen's formulation of micropolar liquid crystals recovers EL nematodynamics, upon assuming that all molecules are uniaxial. In turn, this last constraint enforces the dynamics to neglect the presence of defects (i.e., $\gamma_0=0$), which otherwise would induce variations in the molecule shape.

\subsection{Micropolar theory for uniaxial nematics}

In Eringen's theory, the assumption of uniaxial molecules leads to a microinertia tensor of the form
\begin{equation}
j=J(\mathbf{I}-\bn\otimes\bn),
\label{microtensor}
\end{equation}
which corresponds to $i:=\frac{1}{2}\operatorname{Tr}(j)\mathbf{I}-j=J\bn\otimes\bn$. Here we assume that 
$\|\mathbf{n}\|^2=1$.
Then, this relation transforms Eringen's Lagrangian 
$\ell_2(\bnu,j,\bgamma)$, given in \eqref{reduced_Eringen_Lagr}, to 
\begin{equation}
\begin{aligned}
\ell_2'( \boldsymbol{\nu},  \mathbf{n},\bgamma ):&=\ell_2( \boldsymbol{\nu}, J(\mathbf{I}-\mathbf{n}\otimes\mathbf{n}),\bgamma)\\
&= \frac{J}{2} \int_{ \mathcal{D} }  \| \boldsymbol\nu\times\bn\|^2 \mu -\int_{ \mathcal{D} }\Psi (J(\mathbf{I}-\bn\otimes\bn) , \bgamma ) \mu,
\end{aligned}
\end{equation}
thereby producing the equations \eqref{EL_gamma}, where the initial condition $\gamma_0 = 0$ has not yet been imposed. Since we already know how equations  \eqref{EL_gamma}, with Lagrangian \eqref{SecondLagrangian}, are related to EL nematodynamics, we need to choose a free energy $\Psi$ which equals the Frank energy \eqref{standard_energy}. Thus, we need to prove that there exists $\Psi$ such that
\begin{equation}\label{energyrelations}
\Psi(j,\bgamma)=\Psi(J(\mathbf{I}-\bn\otimes\bn),\bgamma)=F(\bn,\bn\times\bgamma)=F(\bn,\nabla\bn),
\end{equation}
where $F$ is the Frank energy and the last equality follows from the relation $\nabla\bn=\bn\times\bgamma$, which is preserved by the dynamics \eqref{EL_gamma}. Notice that imposing $\nabla\bn=\bn\times\bgamma$ amounts to considering a subsystem of \eqref{EL_gamma}, unless $\gamma_0=0$. The next section shows how an appropriate free energy $\Psi$ can be derived.

\subsection{The expression of the free energy}

This section shows how all terms in the Frank energy \eqref{standard_energy} can be rewritten in terms of the variables $j=J(\mathbf{I}-\bn\otimes\bn)$ and $\bgamma$, the latter being introduced through the invariant relation $\nabla\bn=\bn\times\bgamma$. Thus, the explicit expression for the micropolar free energy $\Psi(j,\gamma)$ of nematic media will be written after computing the micropolar expression for each term in the Frank energy \eqref{standard_energy}. Some equalities are shown in detail in Appendix \ref{Appendix}.

\paragraph{Twist.} Using $\mathbf{n}\otimes\mathbf{n}=\mathbf{I}-{j}/J$, we have
\[
\mathbf{n}\cdot \nabla\times\mathbf{n}=-\mathbf{n}\cdot\boldsymbol{\gamma}(\mathbf{n})+\|\mathbf{n}\|^2\operatorname{Tr}(\boldsymbol{\gamma})=\frac{1}{J}\operatorname{Tr}\left({j}\bgamma\right)=\frac{1}{J}\operatorname{Tr}\left({j}\bgamma^S\right).
\]
where $\boldsymbol{\gamma}_i(\mathbf{n}):=\bgamma_{ia}n_a$, with 
$a$ being the $\mathfrak{so}(3)\simeq\Bbb{R}^3$-index, and 
$\bgamma^S$ denotes the symmetric part of $\bgamma$, i.e., 
$\bgamma^S=\left(\bgamma+\bgamma^T\right)\!/2$, where we see 
$\bgamma$ as a $3\times 3$ matrix with components $\bgamma_{ia}$.

\paragraph{Splay.} We introduce the vector $({\underline{\bgamma}})_b=\epsilon_{abc}\bgamma_{ac}$, defined by the condition $\underline{\boldsymbol{\gamma }} \cdot\mathbf{u}=\operatorname{Tr}(\mathbf{u}\times\boldsymbol{\gamma})$, for all $\mathbf{u}\in\mathbb{R}^3$, where $\mathbf{u}\times\boldsymbol{\gamma}$ is the matrix with components $(\mathbf{u}\times\boldsymbol{\gamma})_{ia}=(\mathbf{u}\times\boldsymbol{\gamma}_i)_a$. We compute
\begin{align*}
\left(\operatorname{div}\mathbf{n}\right)^2&=( \underline{\boldsymbol{\gamma }} \cdot \mathbf{n} )(\underline{\boldsymbol{\gamma }} \cdot \mathbf{n})=\underline{\boldsymbol{\gamma }} \cdot ( \mathbf{n} \otimes \mathbf{n} ) \underline{\boldsymbol{\gamma }}\\
&=\underline{\boldsymbol{\gamma }} \cdot ( \mathbf{I} - {j}/J) \underline{\boldsymbol{\gamma }}= \|\underline{\boldsymbol{\gamma }}\|^2- \frac{1}{J} \underline{\boldsymbol{\gamma }}\cdot{j}\underline{\boldsymbol{\gamma }}\\
&=2\left(\operatorname{Tr}(j)/J-1\right)\operatorname{Tr}\left((\bgamma^A)^2\right)-\frac{4}{J}\operatorname{Tr}\left(j(\bgamma^A)^2\right),
\end{align*} 
where $\bgamma^A$ denotes the skew part of $\bgamma$, i.e. $\bgamma^A=\left(\bgamma-\bgamma^T\right)\!/2$ and where we used the equality $\widehat{\ \underline{\boldsymbol{\gamma}}\ }=-2\boldsymbol{\gamma}^A$. The latter can be shown by noting that we have the equalities $\operatorname{Tr}\left(\widehat{\ \underline{\boldsymbol{\gamma}} \ }\widehat {\mathbf{u}}\right)=-2\underline{\boldsymbol{\gamma}}\cdot\mathbf{u}=-2\operatorname{Tr}(\boldsymbol{\gamma}\widehat{\mathbf{u}})$ for all $\mathbf{u}\in\mathbb{R}^3$.

\paragraph{Bend.} For all $ \mathbf{u} \in \mathbb{R}  ^3 $, we have
\begin{align*}
(\mathbf{n} \times ( \nabla \times \mathbf{n} ) ) \cdot \mathbf{u} 
&=-\nabla _{ \mathbf{n} } \mathbf{n}  \cdot \mathbf{u} 
= -(\mathbf{n} \times \boldsymbol{\gamma}( \mathbf{n} )) \cdot \mathbf{u} 
=-(\mathbf{u} \times \mathbf{n} ) \cdot 
\boldsymbol{\gamma}( \mathbf{n} )\\
&= -\widehat{\mathbf{u}} \mathbf{n}  \cdot 
\boldsymbol{\gamma}( \mathbf{n} ) 
= -\operatorname{Tr}((\widehat{ \mathbf{u}}\mathbf{n})^T 
\boldsymbol{\gamma}\mathbf{n}  )\\
&=\operatorname{Tr} \left((\mathbf{n} \otimes \mathbf{n})
\widehat{ \mathbf{u} } \boldsymbol{\gamma}\right) 
=\operatorname{Tr}\left((\mathbf{I} - {j}/J)
\widehat{ \mathbf{u} } \boldsymbol{\gamma} \right)\\
&=  \operatorname{Tr}( \mathbf{u} \times \boldsymbol{\gamma} )-\frac{1}{J}\operatorname{Tr} (\mathbf{u} \times ( \boldsymbol{ \gamma } {j}))\\
&=\underline{\boldsymbol{\gamma }} \cdot \mathbf{u} - 
\frac{1}{J}\, \underline{\boldsymbol{\gamma }{j}\,} \cdot \mathbf{u},
\end{align*} 
so we get
\[
\mathbf{n} \times ( \nabla \times \mathbf{n} )=\underline{\boldsymbol{\gamma }}- \frac{1}{J}\,\underline{\boldsymbol{\gamma }{j}\,} 
\]
and therefore
\[
\|\mathbf{n} \times ( \nabla \times \mathbf{n} )\|^2=\left\|\frac{1}{J}\,\underline{\boldsymbol{\gamma }{j}} -\underline{\boldsymbol{\gamma }}\right\|^2=-2\operatorname{Tr}\left(\left(\frac{1}{J}(\bgamma j)^A-\bgamma^A\right)^{\!2}\right).
\]
Thus, we proved the following result.
\begin{proposition}{\rm (Eringen \cite{Eringen1993})}
Upon using the relations $\nabla \mathbf{n}= \mathbf{n} \times \bgamma$ and  $j=J(\mathbf{I}-\bn\otimes\bn)$,  the micropolar form of the Frank free energy \eqref{standard_energy} is given by
\begin{multline}\label{FrankMicropolar}
\Psi(j,\bgamma)=\frac{K_2}{J}\operatorname{Tr}(j\bgamma)+ \frac{K_{11}}J\Big(\!\operatorname{Tr}\!\left((\bgamma^A)^2\right)\left(\operatorname{Tr}(j)-J\right)-2\operatorname{Tr}\!\left(j(\bgamma^A)^2\right)\!\Big)
\\
+\frac12\frac{K_{22}}{J^2}\operatorname{Tr}^2(j\bgamma)
-\frac{K_{33}}J\operatorname{Tr}\!\left(\!\left((\bgamma j)^A-J\bgamma^A\right)^{\!2}\right).
\end{multline}
\end{proposition}

\begin{remark}[Surface terms]{\rm Notice that analogous
expressions can be found for certain surface terms that are often added to
the expression \eqref{standard_energy} for the Frank energy; for example, $\operatorname{Tr}\left((\nabla \mathbf{n})^2 \right) = -\gamma_i^h\gamma_j^l \varepsilon_{jhk} \varepsilon_{ilp} \left(\boldsymbol{1} - 
{j}/{J}\right)^{kp}$.
\quad $\blacklozenge$}
\end{remark}
When $j = J(\mathbf{I} - \mathbf{n}\otimes\mathbf{n})$ and the above free energy $\Psi(j,\gamma)$ replaces 
$\Upsilon_2(j, \boldsymbol{\gamma})$ in equations 
\eqref{EP3-B}, the latter are completely equivalent to the Ericksen-Leslie equation \eqref{ELeslie}. This is shown in the next subsection.

\subsection{Recovering Ericksen-Leslie nematodynamics}\label{FromErToEL}
The relations $j=J(\mathbf{I}-\bn\otimes\bn)$ and $\nabla\bn=\bn\times\bgamma$  implied that the Frank energy has a micropolar formulation $\Psi(j,\bgamma)=F(\bn,\bn\times\bgamma)$, to be used in the Lagrangian $\ell_2(\bnu,j,\bgamma)$. Here we restrict to the case $\bgamma_0=0$, so that the equations \eqref{EL_gamma} arising from $\ell'_2$ can now be transformed into a system of the type \eqref{EL}. In this way, the reduced Lagrangian $\ell_2(\bnu,\bn,\bgamma)$ transforms to a functional  $\ell_1'( \boldsymbol{\nu},  \mathbf{n} )$, by using $F(\bn,\bn\times\bgamma)=F(\bn,\nabla\bn)$. Then, the resulting Lagrangian 
\begin{equation}\label{l_1_prime}
\ell_1'( \boldsymbol{\nu},  \mathbf{n} )= \frac{J}{2} \int_{ \mathcal{D} }\left \| \boldsymbol\nu\times\bn \right \|^2 \mu -\int_{ \mathcal{D} }F ( \mathbf{n} , \nabla \mathbf{n} ) \mu\,,
\end{equation}
coincides with \eqref{ELlagrangian1}. 
The procedure outlined in Section \ref{sec:ErLe} (see also Theorem \ref{first_compatibility_thm})
shows that Eringen's micropolar theory recovers Ericksen-Leslie nematodynamics. It is worth emphasizing that the solutions of the Ericksen-Leslie equations 
arising from the Lagrangian $\ell_1'$ still lie on the  zero-level set
$ \mathbf{n}\cdot{\delta \ell _1'  }/{\delta \boldsymbol{ \nu } } =0$,
thereby showing that there is no angular momentum $J\,\mathbf{n}\times\partial_t{\mathbf{n}}$ along the director field $\bn$. Again, this reflects the uniaxial nature of the nematic molecules described by Ericksen-Leslie theory.

Summarizing these considerations and the equivalence between equations \eqref{ELeslie} and 
\eqref{EP_form_EL}, we get the following result.

\begin{theorem}
Under the assumption $j = J(\boldsymbol{1} - 
\mathbf{n} \otimes \mathbf{n})$, the system of equations 
\eqref{EP3-B} with $\Upsilon_2(j, \boldsymbol{\gamma}) = 
\Psi(j, \boldsymbol{\gamma})$ given in 
\eqref{FrankMicropolar}, are equivalent to the Ericksen-Leslie 
equations \eqref{ELeslie}.
\end{theorem}
\begin{remark}[The one-constant approximation]\label{rem:one_constant}\rm 
Notice that, for the one constant approximation, $K_2=0$ and $K_{11}=K_{22}=K_{33}=:K$ and the Frank free energy $F(\bn,\nabla\bn)=(K/2)\!\int_{\mathcal{D}}\|\nabla\bn\|^2\mu$ yields expression  \eqref{FrankMicropolar} in the form
\[
\Psi(j,\bgamma)=\frac{K}{2J}(\bgamma_i\cdot j\bgamma_i)
\,.
\]
Then, the resulting equations of Eringen's micropolar theory are obtained upon replacing $\Upsilon_2(j,\bgamma)=\Psi(j,\bgamma)$ in \eqref{EP3-B}. Eventually, one obtains the equations of motion:
\begin{equation}
\left\{
\begin{array}{l}
\vspace{0.2cm}\displaystyle J\big(j\partial_t{\bnu}-
j\bnu\times\bnu\big)=-K\big(j{\partial_{x_i}\bgamma_i}+j\bgamma_i\times\bgamma_i
\big)
\\
\vspace{0.2cm}\displaystyle \partial_t j+[j,\hat{\boldsymbol{\nu}}]=0,\\
\displaystyle \partial _t \bgamma_i +  \bgamma_i \times \boldsymbol{\nu}+ {\partial_{x_i}} \boldsymbol{\nu}=0,\quad{\bgamma}_0=0\,,
\end{array}\right.
\label{oneconstant}
\end{equation}
where we have used the relations \eqref{objectivity} and $\nabla j=\left[j,\widehat{\gamma}\right]$. The above equations can also be expressed in terms of the director field $\bn$, upon using the relation \eqref{microtensor}. It is interesting to observe that the first equation in \eqref{oneconstant} can be rewritten in the particularly suggestive form
\[
J\partial_t\!\left({j\bnu}\right)+K\partial_{x_i\!}\!\left({j\bgamma_i}\right)=0
\,,
\]
which is then accompanied by  $\nabla j=\left[j,\widehat{\gamma}\right]$ as well as the second and third equations in \eqref{oneconstant}.
\quad $\blacklozenge$
\end{remark}

\subsection{Remarks on biaxial nematics}

The case of biaxial liquid crystals offers a good opportunity to express all variables in terms of the order parameter quantities. For example, the well known expression \cite{KlLa2003}
\[
j=J_1\left(\mathbf{I}-\bn\otimes\bn\right)+J_2\left(\bm\otimes\bm-(\bn\times\bm)\otimes(\bn\times\bm)\right)
\]
of the (microinertia) tensor order parameter in terms of the two directors $\bn$ and $\bm$ (with $\bn\cdot\bm=0$), was shown in \cite{GBTr2010} to remain invariant in the absence of disclinations, i.e., when $\bgamma_0=0$ and  $\hat{\bgamma}=-(\nabla\chi)\chi^{-1} $. In this case, the above expression of the tensor order parameter yields a Lagrangian of the type $\ell(\bnu,\bn,\bm)$ where $\widehat\bnu=\dot\chi\chi^{-1}$ and $(\bn,\bm)=(\chi\bn_0,\chi\bm_0)$ (see \cite{GBTr2010}). Then, upon choosing the initial conditions $\bn_0=\mathbf{e}_3$ and $\bm_0=\mathbf{e}_2$, the orthogonal matrix $\chi$ can be expressed as
\begin{displaymath}
\chi=
\left(\bm\times\bn\quad\bm\quad\bn\right)
\end{displaymath}
Then, by using the orthogonality of the rows and columns of $\chi$ and $\chi^T$, one has
\begin{align}
&\bnu=\frac{1}{2} \left( \bn\times\partial_t\bn+\bm\times\partial_t\bm+(\bm\times\bn)\times\partial_t(\bm\times\bn)\right) 
\\
&\bgamma=\frac{1}{2} \left( \nabla\bn\times\bn+\nabla\bm\times\bm+\nabla(\bm\times\bn)\times(\bm\times\bn)\right) 
\label{gamma-biaxials}
\end{align}
so that all dynamical variables are expressed explicitly in terms of the two directors. Notice that if $J_2=0$, the case of uniaxial molecules treated previously in this section, prevents the potential $\bgamma$ to be expressed uniquely in terms of the director $\bn$. Indeed, setting $\bm=0$ is evidently forbidden by the orthogonality property of $\chi\in SO(3)$. This fact is particularly interesting because it contrasts with Eringen's definition in \cite{Eringen1993, Eringen1997}, which suffers from not being gauge invariant.

On the other hand, the dynamics of biaxial nematics in the absence of disclinations is completely equivalent to the Euler-Lagrange dynamics arising from Hamilton's principle $\delta\int_{t_1}^{t_2}\!\mathcal{L}(\chi,\dot\chi)\,\de t=0$ on the total space $T\mathcal{F}(\Bbb{R}^3,SO(3))$ since the two directors completely determine the rotation matrix $\chi$, which then identifies by itself all degrees of freedom of the system. This is due to the fact that for biaxial molecules, the rotational symmetry is completely broken and thus reduction theory returns the total space $T\mathcal{F}(\Bbb{R}^3,SO(3))$.

Then, in this case one observes that the process illustrated in Section \ref{sec_first_reduction} can still be implemented according to
\begin{align*}
\mathcal{L}(\chi,\dot\chi)=&\int_\mathcal{D}\mathscr{L}(\chi,\dot\chi,\chi\bn_0,\chi\bm_0,\nabla\chi\bn_0,\nabla\chi\bm_0)
\\
=&\int_\mathcal{D}\mathscr{L}(\dot\chi\chi^{-1},\bn,\bm,\nabla\bn,\nabla\bm)=\ell_1(\bnu,\bn,\bm)
\,.
\end{align*}
However, the directors $(\bn,\bm)$ completely determine the rotation matrix $\chi$, so that $\ell_1(\bnu,\bn,\bm)=\ell(\bnu,\chi)$. This means that the reduced space coincides with the total space $T\mathcal{F}(\Bbb{R}^3,SO(3))$, as it can be seen by applying the inverse trivialization $(\widehat\bnu,\chi)=(\dot\chi\chi^{-1},\chi)\mapsto(\dot\chi,\chi)$. Notice that when inertial effects are neglected, which amounts to enforcing $\left(\partial_t+\bnu\times\right)\delta \ell/\delta\bnu=0$, the equations resulting from this approach (see \cite{GBTr2010}) coincide with those found by \cite{VoKa1981}.

 The variable $\widehat\bgamma=\nabla\chi\chi^{-1}$ can be also introduced for physical purposes by following the procedure outlined in Section \ref{sec_second_reduction}. An example of how this quantity is used in condensed matter media is provided by frustrated spin glasses; see \cite{Volovick1980,GBTr2011} and references therein.

When disclinations are present in biaxial nematic media, then the gauge-invariant potential $\widehat\bgamma=\chi\widehat\bgamma_0\chi^{-1}+\chi\nabla\chi^{-1}$ appears as an extra variable in the system, since in this case $\bgamma_0\neq 0$ so that $\bgamma$ possesses its own evolution. Again, this situation fits in the description given by equations \eqref{EP_continuum2}:  the order parameter $n$ coincides with the two directors $n=(\bn,\bm)$ and the Lagrangian is of the type $\ell_2(\bnu,\bn,\bm,\bgamma)$, so that the momentum map in \eqref{EP_continuum2} is given by
\[
{\bf J}\!\left(\frac{\delta \ell_2}{\delta \bn},\frac{\delta \ell_2}{\delta \bm}\right)=\bn\times\frac{\delta \ell_2}{\delta \bn}+\bm\times\frac{\delta \ell_2}{\delta \bm}\,.
\]
Notice that in this case the relations $(\nabla\bn_0,\nabla\bm_0)=(\bn_0\times\bgamma_0,\bm_0\times\bgamma_0)$ prevent establishing a unique correspondence between the directors $(\bn,\bm)$ and the rotation matrix $\chi$, so that the potential $\bgamma$ can no longer be expressed explicitly in terms of the directors, as it was done in equation \eqref{gamma-biaxials}.

\section{Two reductions for the Lhuillier-Rey theory}

The  Lhuillier-Rey theory is an alternative description for liquid crystals of molecules with variable shape \cite{LR}. In this description, the order parameter field is given by two components: the microinertia tensor $j$ and the director field $\bn$. In accordance with Section \ref{sec:GeneralCase}, we continue assuming that $\nabla \mathbf{n}_0 = 0$ and $\nabla j_0 = 0$. Because of the coexistence of these two order parameters, one can keep the expression of the Frank free energy \eqref{standard_energy} in the Lagrangian while allowing for a variable molecular shape, represented by the microinertia tensor appearing in the kinetic energy. The unreduced Lhuillier-Rey Lagrangian is 
\begin{align}\label{LR-UnreducedLagrangian}
\mathcal{L} ( \chi , \dot \chi )
&=\int_{ \mathcal{D} }\mathscr{L} ( \chi , \dot \chi,\chi\bn_0,\chi j_0\chi^{-1},{\nabla}(\chi\mathbf{n}_0) ) \mu
\nonumber \\
&=\frac{1}{2}\int_{ \mathcal{D} } \operatorname{Tr}\left( (i_0 \chi ^{-1} \dot \chi ) ^T \chi ^{-1} \dot \chi  \right)  \mu -\int_{ \mathcal{D} }F( \chi \mathbf{n} _0 , {\nabla} ( \chi \mathbf{n} _0 )) \mu\,.
\end{align}
Again, this Lagrangian possesses two reductions, which are produced by the invariance properties
\begin{align}
\mathcal{L} ( \chi , \dot \chi )
&=\int_{ \mathcal{D} }\mathscr{L} ( \chi , \dot \chi,\chi\bn_0,\chi j_0\chi^{-1},{\nabla}(\chi\mathbf{n}_0)  ) \mu
\nonumber \\
&=
\int_{ \mathcal{D} }\mathscr{L} ( {\dot \chi}\chi^{-1},\chi\bn_0,\chi j_0\chi^{-1},{\nabla}(\chi\mathbf{n}_0)  ) \mu=\ell_1(\bnu,j,\bn)
\end{align}
and
\begin{align}
\mathcal{L} ( \chi , \dot \chi )
&=\int_{ \mathcal{D} }\mathscr{L} ( \chi , \dot \chi,\chi\bn_0,\chi j_0\chi^{-1} ) \mu
\nonumber \\
&=
\int_{ \mathcal{D} }\mathscr{L} ( {\dot \chi}\chi^{-1},\chi\bn_0,\chi j_0\chi^{-1},-({\nabla}\chi)\chi^{-1} ) \mu=\ell_2(\bnu,j,\bn,\gamma),
\end{align}
where the second invariance arises naturally when $\bn_0$ is constant in space, since one has ${\nabla}\bn=({\nabla}\chi)\chi^{-1}\bn$.

The first invariance yields the Euler-Poincar\'e form of the Lhuiller-Rey equations
\begin{equation}\label{EP-LR1}
\left\{
\begin{array}{l}
\vspace{0.2cm}
\displaystyle
\frac{\partial}{\partial t}\frac{\delta \ell _1 }{\delta \boldsymbol{\nu}}=
\boldsymbol{\nu}\times 
\frac{\delta \ell _1 }{\delta \boldsymbol{\nu}}-\!\overrightarrow{\,\left[\frac{\delta \ell _1}{\delta j}, j\right]\,}+
\mathbf{n} \times \frac{\delta \ell_1}{\delta \mathbf{n}}\\
\vspace{0.2cm}\displaystyle
\partial_t j+[j,\hat{\boldsymbol{\nu}}]=0\\
\displaystyle \partial_ t \mathbf{n} + \mathbf{n} \times \boldsymbol{\nu }=0\,,
\end{array}\right.
\end{equation}
while reduction for the second group action yields the equivalent set of equations
\begin{equation}\label{EP-LR2}
\left\{
\begin{array}{l}
\vspace{0.2cm}\displaystyle\frac{\partial}{\partial t}\frac{\delta \ell _2 }{\delta \boldsymbol{\nu}}=\boldsymbol{\nu}\times \frac{\delta \ell _2}{\delta \boldsymbol{\nu}}-\!\overrightarrow{\,\left[\frac{\delta \ell _2}{\delta j}, j\right]\,}+\mathbf{n} \times \frac{\delta \ell _2 }{\delta \mathbf{n} } +\operatorname{div}^\bgamma\frac{\delta\ell_2}{\delta\bgamma} \\
\vspace{0.2cm}\displaystyle\partial_t j+[j,\hat{\boldsymbol{\nu}}]=0\\
\vspace{0.2cm}\displaystyle \partial_ t \mathbf{n} + \mathbf{n} \times \boldsymbol{\nu }=0\\
\vspace{0.2cm}\displaystyle \partial _t \bgamma +  \bgamma \times \boldsymbol{\nu}+ {\nabla} \boldsymbol{\nu}=0,\quad{\bgamma}_0=0,
\end{array}\right.
\end{equation}
where we used the relation $F(\mathbf{n},\nabla\bn)=F(\mathbf{n,n}\times\boldsymbol\gamma)$ (see Appendix \ref{Appendix}).
The Lagrangian \eqref{LR-UnreducedLagrangian} yields
\begin{equation}
\ell_1(\bnu,j,\bn)=\frac{J}{2} \int_{ \mathcal{D} }\left \| j\boldsymbol\nu\right \|^2 \mu -\int_{ \mathcal{D} }F ( \mathbf{n} , \nabla \mathbf{n} ) \mu
\end{equation}
and
\begin{equation}
\ell_2(\bnu,j,\bn,\bgamma)=\frac{J}{2} \int_{ \mathcal{D} }\left \| j\boldsymbol\nu\right \|^2 \mu -\int_{ \mathcal{D} }F ( \mathbf{n} ,  \mathbf{n}\times\bgamma ) \mu
\,,
\end{equation}
so that the systems \eqref{EP-LR1}-\eqref{EP-LR1} become 
\begin{equation}\label{EP-LR1-A}
\left\{
\begin{array}{l}
\vspace{0.2cm}\displaystyle 
j\partial_ t{\bnu}=j\bnu\times\bnu-\mathbf{n} \times \mathbf{h}\\
\vspace{0.2cm}\displaystyle
\partial_t j+[j,\hat{\boldsymbol{\nu}}]=0\\
\vspace{0.2cm}\displaystyle \partial_ t \mathbf{n} + \mathbf{n} \times \boldsymbol{\nu }=0\,,
\end{array}\right.
\end{equation}
(where the molecular field $\mathbf{h}$ is given in \eqref{molecularfield1}) and
\begin{equation}\label{EP-LR1-B}
\left\{
\begin{array}{l}
\vspace{0.2cm}\displaystyle 
j\frac{\partial\bnu}{\partial t}=
j\bnu\times\bnu-\frac{\partial}{\partial x^i}\frac{\partial \Phi}{\partial\bgamma_i}-{\boldsymbol\gamma_i}  \times\frac{\partial \Phi}{\partial\bgamma_i}-\bn\times\frac{\partial \Phi}{\partial \bn}\\
\vspace{0.2cm}\displaystyle
\partial_t j+[j,\hat{\boldsymbol{\nu}}]=0\\
\vspace{0.2cm}\displaystyle \partial_ t \mathbf{n} + \mathbf{n} \times \boldsymbol{\nu }=0,\\
\displaystyle \partial _t \bgamma +  \bgamma \times \boldsymbol{\nu}+ {\nabla} \boldsymbol{\nu}=0,\quad{\bgamma}_0=0\,,
\end{array}\right.
\end{equation}
where $\Phi(\bn,\bgamma)$ is given as in \eqref{Frank-variant}, as it arises from the relation 
$\nabla\bn=\bn\times\bgamma$. In view of the identity
\eqref{identity_useful_EL}, we get the following result.

\begin{theorem}
The systems of 
equations \eqref{EP-LR1-A} and \eqref{EP-LR1-B} are equivalent.
\end{theorem}

Notice that, the relations $({\nabla} j,{\nabla}\mathbf{n})=(\left[j,\widehat{\bgamma}\right],\mathbf{n}\times\boldsymbol\gamma)$ determine an invariant subsystem, regardless of the initial conditions $(j_0,\bn_0,\bgamma_0)$. Indeed, the possibility of an inhomogeneous initial condition $\bgamma_0\neq0$ extends the Lhuiller-Rey theory to account for disclination dynamics. Upon taking $\mathcal{O}=SO(3)$ and $M=\mathrm{Sym}(3)\times S^2$, it is easy to see that the two Lhuiller-Rey formulations \eqref{EP-LR1} and \eqref{EP-LR2} follow directly from applying the general theory of Section \ref{sec:GeneralCase} to the Lagrangian \eqref{LR-UnreducedLagrangian}.

It is interesting to notice that similar arguments to those in Section \ref{FromErToEL} show immediately how Lhuiller-Rey theory recovers Ericksen-Leslie nematodynamics. Indeed, while the relation ${\nabla}\mathbf{n}=\mathbf{n}\times\boldsymbol\gamma$ can be used to transform $\ell_2$ into $\ell_1$, the initial condition
\[
j _0 = J(\mathbf{I}- \mathbf{n} _0 \otimes \mathbf{n} _0 )
\]
readily produces the Lagrangian \eqref{l_1_prime}, as was shown in Section \ref{FromErToEL}, to recover the Ericksen-Leslie equations.

\section{Flowing liquid crystals}

This section extends the previous discussions to the case of flowing liquid crystals. In this case, upon denoting the fluid flow by the diffeomorphism $\eta\in\mathrm{Diff}(\mathcal{D})$,  the gradient of the director field $\mathbf{n}=(\chi\,\mathbf{n}_0)\circ\eta^{-1}=\eta_*(\chi\,\mathbf{n}_0)$ is computed to be
\[
{\nabla}\mathbf{n}=
\eta_*\big(\nabla\chi\,\mathbf{n}_0\big)=\eta_*\big(\!\left(\nabla\chi\chi^{-1}\right)\chi\mathbf{n}_0\big)=\left(\eta_*\!\left(\nabla\chi\chi^{-1}\right)\right)\left(\eta_*(\chi\mathbf{n}_0)\right)=:\mathbf{n} \times\boldsymbol\gamma\,,
\]
where we have used standard properties of the push-forward $\eta_*$  and we have defined 
\[
\widehat{\bgamma}=-\eta_*\!\left(\nabla\chi\chi^{-1}\right).
\]
A similar argument actually holds for any order parameter space $M$ that is acted upon by a Lie group $\mathcal{O}$. In this more general case, the previous relation reads as \eqref{GenFormulaGradient}. This suggests that all the considerations in Section \ref{sec:GeneralCase} have a natural correspondent in the case of liquid crystal flows. 

\subsection{Euler-Poincar\'e and affine Euler-Poincar\'e reductions}
Upon restricting to incompressible fluid flows for convenience, the general form of the Lagrangian for flowing uniaxial liquid crystals is
\[
\mathcal{L}(\eta,\dot\eta,\chi,\dot\chi)=
\int_\mathcal{D}\mathscr{L}(\eta,\dot\eta,\chi,\dot\chi,
(\chi_0)\circ\eta^{-1},{\nabla}\!\left(
(\chi n_0)\circ\eta^{-1}\right)\,\mu\,,
\]
so that $\mathcal{L}$ is a functional of the type
\[
\mathcal{L}:T\big(\mathrm{Diff}_{\rm vol}(\mathcal{D})\,\circledS\,\mathcal{F(D,O})\big)\to\Bbb{R}
\]
where the semidirect product structure reflects the fact that the microstructure variable $\chi\in\mathcal{F(D,O})$ is pulled around the domain $\mathcal{D}$ by the fluid flow $\eta\in\mathrm{Diff}_{\rm vol}(\mathcal{D})$ \cite{GBRa2009}. 
More precisely, the Lagrangian of a flowing uniaxial liquid crystal is 	
\[
\mathcal{L}_{n_0}(\eta,\dot\eta,\chi,\dot\chi)=\frac{1}{2}\int_\mathcal{D}\|\dot\eta\|^2\mu+\frac{1}{2}\,J\!\int_\mathcal{D}\|\dot\chi\|^2\mu-\int_\mathcal{D}F\left((\chi n_0)\circ\eta^{-1},{\nabla} ((\chi n_0)\circ\eta^{-1})\right)\mu\,,
\]
where $F$ is the Frank free energy. Again, the fixed order parameter $n_0$ appears in the Lagrangian density $\mathscr{L}$ through both $\chi n_0$ and its differential ${\nabla}(\chi n_0)$. Then, the Lagrangian $\mathcal{L}$ possesses the following invariance properties:
\begin{align}\nonumber
\mathcal{L}(\chi,\dot\chi)=&
\int_\mathcal{D}\mathscr{L}(\eta,\dot\eta,\chi,\dot\chi,\chi n_0,{\nabla}(\chi n_0))\,\mu
\\\nonumber
=&\int_\mathcal{D}\mathscr{L}(\dot\eta\circ\eta^{-1},(\dot\chi\chi^{-1})\circ\eta^{-1},(\chi n_0)\circ\eta^{-1},{\nabla}((\chi  n_0)\circ\eta^{-1}))\,\mu
\\
=&\!:\int_\mathcal{D}\mathscr{L}(\mathbf{u},\nu,n,\nabla n)\,\mu
=\ell_1(\mathbf{u},\nu,n),
\label{invariance1-bis}
\end{align}
and, if $n_0$ is constant in space,
\begin{align}\nonumber
\mathcal{L}(\chi,\dot\chi)=&
\int_\mathcal{D}\mathscr{L}(\eta,\dot\eta,\chi,\dot\chi,\chi n_0,{\nabla}(\chi n_0))\,\mu
\\\nonumber
=&\int_\mathcal{D}\mathscr{L}(\dot\eta\circ\eta^{-1},(\dot\chi\chi^{-1})\circ\eta^{-1},(\chi n_0)\circ\eta^{-1},(\eta_*(({\nabla}\chi)\,\chi^{-1}))((\chi n_0)\circ\eta^{-1}) )\,\mu
\\
=&\!:\int_\mathcal{D}\mathscr{L}(\mathbf{u},\nu,n,-\gamma_M(n))\,\mu
=\ell_2(\mathbf{u},\nu,n,\gamma)\,,
\label{invariance2-bis-fluid}
\end{align}
where one defines $\mathbf{u}:=\dot\eta\circ\eta^{-1}\in\mathfrak{X}_{\rm div}(\mathcal{D})$ (the space of divergence
free vector fields tangent to the boundary of $\mathcal{D}$),  $\nu:=(\dot{\chi}\chi^{-1})\circ\eta^{-1}\in\mathcal{F}(\mathcal{D},\mathfrak{o})$, $n:=(\chi n_0)\circ\eta^{-1}\in\mathcal{F}(\mathcal{D},M)$, and $\widehat\gamma:=-\eta_*(({\nabla}\chi)\,\chi^{-1})\in\Omega^1(\mathcal{D},\mathfrak{o})$.

We notice that all the symmetry breaking arguments from Section \ref{sec:GeneralCase} transfer to this case without essential changes. However, in the present case all variables are acted upon by the diffeomorphism $\eta\in\mathrm{Diff}_{\rm vol}(\mathcal{D})$, which affects the way variations are taken in the variational principles
\[
\delta\!\int_{t_0}^{t_1}\ell_1(\mathbf{u},\nu,n)\,\mathrm{d}t=0\,,
\qquad
\delta\!\int_{t_0}^{t_1}\ell_2(\mathbf{u},\nu,n,\gamma)\,\mathrm{d}t=0\,.
\]
Indeed, upon making use of the Lie derivative notation $\boldsymbol{\pounds}$, the variations
\begin{align*}
\delta \mathbf{u}&=\partial_t \mathbf{w}+\boldsymbol{\pounds}_\mathbf{u} \mathbf{w}
\\
\delta \nu&=\partial_t \omega -\boldsymbol{\pounds}_\mathbf{w} \nu+\boldsymbol{\pounds}_\mathbf{u} \omega+[\omega,\nu]
\\
\delta n&=-\boldsymbol{\pounds}_\mathbf{w} n+\omega_M(n)
\\
\delta\gamma&=-\boldsymbol{\pounds}_\mathbf{w} \gamma-{\nabla}^\gamma\omega
\end{align*}
lead to the replacement all time derivatives in the equations 
\eqref{EP_continuum} and \eqref{EP_continuum2} for motionless 
media by appropriate material derivatives, as is usual in the 
Euler-Poincar\'e theory of fluid flows \cite{HoMaRa1998}.
Here, $\mathbf{w}:=\delta \eta \circ 
\eta^{-1}$ is a path in $\mathfrak{X}_{\rm div}(M)$ and 
$\omega : = (\delta\chi) \chi^{-1}$
is a path in $\mathcal{F}(\mathcal{D},\mathfrak{o})$, 
both vanishing at $t_0$ and $t_1$.
Then, the resulting equations are
\begin{equation}\label{EP_fluid1}
\left\{
\begin{array}{l}
\vspace{0.2cm}\displaystyle
\frac{D}{Dt}\frac{\delta \ell_1}{\delta \mathbf{u}}
+ \left\langle\frac{\delta \ell_1}{\delta \nu},
{\nabla}\nu\right\rangle +
\left\langle\frac{\delta \ell_1}{\delta n}, 
\nabla n\right\rangle=-\nabla p, 
\quad\operatorname{div}\mathbf{u} = 0,
\\
\vspace{0.2cm}\displaystyle
\frac{D}{D t}\frac{\delta
\ell_1}{\delta\nu}+\operatorname{ad}^*_\nu\frac{\delta \ell_1}{\delta\nu}=\mathbf{J}\!\left(\frac{\delta
\ell_1}{\delta n}\right),
\quad 
\frac{Dn}{Dt}=\nu_M( n)\,,
\end{array}\right.
\end{equation}
and
\begin{equation}\label{EP_fluid2}
\!\!\!\!
\left\{
\begin{array}{l}
\vspace{0.2cm}\displaystyle
\frac{D}{Dt}\frac{\delta \ell_2}{\delta \mathbf{u}}
+\left\langle\frac{\delta \ell_2}{\delta \nu},
\nabla\nu\right\rangle +
\left\langle \frac{\delta \ell_2}{\delta n}, 
\nabla n\right\rangle - \frac{\delta \ell_2}{\delta \gamma}\diamond\gamma=-\nabla p, \quad
\operatorname{div}\mathbf{u} = 0,
\\
\vspace{0.2cm}\displaystyle
\frac{D}{D t}\frac{\delta
\ell_2}{\delta\nu}+\operatorname{ad}^*_\nu\frac{\delta \ell_2}{\delta\nu}=\mathbf{J}\!\left(\frac{\delta
\ell_2}{\delta n}\right)+\operatorname{div}^{\gamma\!}\left(\frac{\delta \ell_2}{\delta \gamma}\right),
\quad 
\left(\frac{Dn}{Dt},\frac{D\gamma}{Dt}\right)=\left(\nu_M (n),-{\nabla}^\gamma\nu\right),
\end{array}\right.
\end{equation}
where the material time derivative $D/Dt$ is expressed in terms of the Lie derivative operator $\boldsymbol{\pounds}_{\mathbf{u}}$  as
\[
\frac{D}{Dt}=\frac{\partial}{\partial t}+\boldsymbol{\pounds}_{\mathbf{u}}
\,,
\]
while the diamond operator is defined by
\[
\left\langle\kappa\diamond\gamma,\mathbf{u}\right\rangle:=-\left\langle\kappa,\boldsymbol{\pounds}_\mathbf{u}\gamma\right\rangle, \quad \kappa \in \mathfrak{X}(\mathcal{D},\mathfrak{o}^\ast), \quad \gamma\in\Omega^1(\mathcal{D}, \mathfrak{o})
\]
and its expression in local coordinates is $(\kappa\diamond \gamma)_h = -\kappa^j_a \partial_h\gamma_j^a + \partial_j(\kappa^j_a \gamma^a_h)$. The pressure $p$ in
\eqref{EP_fluid1} and \eqref{EP_fluid2} and in all subsequent sections is determined like in the standard ideal incompressible homogeneous Euler equations. This means that  
$p$ is the solution to
the Neumann problem (up to a constant) obtained by requiring that its negative Laplacian equals the divergence of the left hand side of the equation in which $p$ appears and the
boundary condition is given by the negative normal derivative
of $p$ equal to the inner product of the left hand side with 
the unit normal to the boundary. Of course, the divergences are taken here as one-forms on $\mathcal{D}$.

It remains to explain the terms involving a pairing in both
systems \eqref{EP_fluid1}, \eqref{EP_fluid2}. Since $\nu \in 
\mathcal{F}(\mathcal{D}, \mathfrak{o})$, it follows that its
differential  $\nabla \nu: T\mathcal{D} \rightarrow \mathfrak{o}$ and $\delta \ell_1/ \delta\nu \in 
\mathcal{F}(\mathcal{D}, \mathfrak{o}^\ast)$. Thus, 
$\left\langle\frac{\delta \ell_1}{\delta \nu}, \nabla \nu
\right\rangle\in \Omega ^1(\mathcal{D})$ 
is defined by 
\[
\left\langle\frac{\delta \ell_1}{\delta \nu}, 
\nabla \nu \right\rangle (v_x):=
\left\langle\frac{\delta \ell_1}{\delta \nu}(x), 
\nabla \nu(v_x) \right\rangle, 
\]
for any $v_x \in T_x \mathcal{D}$, and the pairing in the
right hand side is the duality pairing $\left\langle\,, 
\right\rangle: \mathfrak{o}^\ast \times \mathfrak{o} 
\rightarrow  \mathbb{R}$.

The understanding of the terms $\left\langle 
\frac{\delta \ell_i}{\delta n}, \nabla n\right\rangle
\in\Omega^1(\mathcal{D})$, $i=1,2$, is more involved and relies
on the observation that $\nabla n: T\mathcal{D} \rightarrow 
TM$ can be thought of as an element of 
$\Omega^1\left( \mathcal{D}, n^\ast(TM)\right)$, the vector
space of $n^\ast(TM)$-valued one-forms on $\mathcal{D}$. Here
$n^\ast(TM):= \left\{(x, u_{n(x)}) \mid x \in \mathcal{D},\;
u_{n(x)}\in T_{n(x)}M \right\} \rightarrow \mathcal{D}$ is
the pull back vector bundle over $\mathcal{D}$ of the tangent
bundle $TM \rightarrow M$ by the map $n: \mathcal{D} \rightarrow M$; recall that the fiber of this pull back
bundle at $x \in \mathcal{D}$ is $T_{n(x)}M$. Consequently,
$\frac{\delta \ell_i}{ \delta n} \in \mathcal{F}\left( \mathcal{D}, n^\ast(T^\ast M) \right)$ and hence
\[
\left\langle \frac{\delta \ell_i}{\delta n}, 
\nabla n\right\rangle(v_x): = 
\left\langle \frac{\delta \ell_i}{\delta n}(x), 
\nabla n(v_x)\right\rangle
\]
for all $v_x \in T_x \mathcal{D}$, where on the right
hand side we use the duality pairing 
$\left\langle\,, \right\rangle: T^\ast_{n(x)}M \times 
T_{n(x)}M \rightarrow \mathbb{R}$.

\begin{remark}[Noether's Theorems] {\rm
Noether's theorems \eqref{Noether1} and \eqref{Noether2-bis} become
\begin{align*}
&\frac{D}{Dt}\,i^*\!\left(\operatorname{Ad}^*_{\chi\circ \eta ^{-1}}\frac{\delta \ell_1}{\delta \nu}\right) = 0\quad\text{or, equivalently,}\quad 
\frac{{\rm d}}{{\rm d}t} i^*\!\left(\operatorname{Ad}^*_{\chi}\left( \frac{\delta \ell_1}{\delta \nu}\circ \eta \right) \right) =0
\quad\text{and}\\ 
&\frac{{\rm d}}{{\rm d}t}\,j^*\!\left( \int_\mathcal{D\!}\left(\operatorname{Ad}^*_{\chi\circ \eta ^{-1}}\frac{\delta\ell_2}{\delta\nu}\right)\mu\right)=0,
\end{align*}
where $i^*$ and $j^*$ are the duals of the Lie algebra inclusions 
$i:\mathcal{F(D},\mathfrak{o}_{n_0}) = \mathcal{F}(\mathcal{D},\mathfrak{o})_{n _0} \hookrightarrow \mathcal{F(D},\mathfrak{o})$
and $j: \mathfrak{o}_{n_0} \hookrightarrow \mathfrak{o}$, respectively.
Note that in the first Noether theorem, the left hand side formulation 
is an Eulerian quantity, whereas the right hand
side is a Lagrangian quantity.

The first Noether theorem follows from the expression of the momentum 
map
\[
\mathbf{J}:T^*(\operatorname{Diff}_{\rm vol}(\mathcal{D}) \,\circledS\, 
\mathcal{F}(\mathcal{D}, \mathcal{O}))\rightarrow 
\mathcal{F}(\mathcal{D},\mathfrak{o} )_{n _0} ^\ast, \quad 
\mathbf{J}  ( \alpha _ \eta , \kappa _\chi )= 
i ^\ast ( \chi ^{-1} \kappa _{ \chi }),
\]
together with the equality $ \kappa _{ \chi } \chi ^{-1} = \frac{\delta \ell _1 }{\delta \nu }\circ \eta  $, which yield
\[
\mathbf{J}  ( \alpha _ \eta , \kappa _\chi ) = i ^\ast ( \chi ^{-1} \kappa _{ \chi })= i ^\ast \left( \operatorname{Ad}^*_{\chi\circ \eta ^{-1}}\frac{\delta \ell_1}{\delta \nu} \right) \circ \eta.
\]
One obtains the second Noether theorem in a similar way.}
\end{remark}

\begin{remark}[From incompressible to compressible flows]{\rm 
Although in this section we focussed on incompressible fluid flows, extending to the compressible case requires no additional argument than those already present in \cite{GBRa2009}. Then, the mass density appears as an additional parameter in the Lagrangian $\mathcal{L}$, whose symmetry properties \eqref{invariance1} and \eqref{invariance2} are accompanied by the relation $\rho_0=\eta_*\rho$ for the mass density $\rho$. This is also explained in \cite{HoMaRa1998}. The dynamics of compressible liquid crystal dynamics can thus be obtained by simply combining the above results with standard principles in Euler-Poincar\'e theory.}
\end{remark}

\subsection{Micropolar liquid crystals and Ericksen-Leslie theory}
This section extends the results of \S\ref{FromErToEL} to account for fluid motion. In the case of micropolar liquid crystals, the order parameter field is $j\in \mathrm{Sym}(3)$ and the Euler-Poincar\'e Lagrangians arising from the two different reductions read
\begin{align*}
\ell_1(\mathbf{u}, \boldsymbol{\nu}, j)=&\frac{1}{2} \int_{ \mathcal{D} } \|\mathbf{u}\|^2   \mu+
\frac{1}{2} \int_{ \mathcal{D} } (j  \boldsymbol{\nu}) \cdot  \boldsymbol{\nu} \mu -\int_{ \mathcal{D} }
\Upsilon_1( j , \nabla j ) \mu,
\end{align*}
and
\begin{align}\label{EringenFluidLagrangian}
\ell_2( \mathbf{u},\boldsymbol{\nu},  j,\bgamma )=&\frac{1}{2} \int_{ \mathcal{D} } \|\mathbf{u}\|^2   \mu+ \frac{1}{2} \int_{ \mathcal{D} } (j  \boldsymbol{\nu}) \cdot  \boldsymbol{\nu} \mu-\int_{ \mathcal{D} }
\Upsilon_2(j , \bgamma ) \mu\,,
\end{align}
where compatibility of the two reductions enforces the energy relations 
\[
\Upsilon_1(j , \nabla j )=\Upsilon_1( j , [j,\widehat{\bgamma}] )=\Upsilon_2(j,\bgamma)\,.
\]
Then, upon specializing to uniaxial nematics, the microinertia tensor is given by
\[
j=J(\mathbf{I}-\mathbf{n}\otimes\mathbf{n})\,,
\]
and the corresponding free energy must transform as
\begin{equation}
\label{all_potential_relations}
\Upsilon_2(j,\bgamma) = 
\Psi(j,\bgamma)=\Psi(J(\mathbf{I} - 
\mathbf{n}\otimes\mathbf{n}),\bgamma)=
F(\mathbf{n},\mathbf{n}\times\bgamma)=F(\mathbf{n},\nabla\mathbf{n})\,,
\end{equation}
with the expression of $\Psi(j, \boldsymbol{\gamma})$ given in \eqref{FrankMicropolar}. 
In conclusion, the Lagrangian $\ell_2$ is transformed to
\[
\ell_1'( \mathbf{u},\boldsymbol{\nu},  \mathbf{n} )=\frac{1}{2} \int_{ \mathcal{D} } \|\mathbf{u}\|^2   \mu+ \frac{1}{2} \,J\int_{ \mathcal{D} }   \|\boldsymbol{\nu}\times\mathbf{n}\|^2 \mu-\int_{ \mathcal{D} }F(\mathbf{n},\nabla\mathbf{n})\mu\,,
\]
which in turn produces the equations \eqref{EP_fluid1} in the form
\begin{equation}\label{EP_EL_fluid}
\left\{
\begin{array}{l}
\vspace{0.2cm}\displaystyle
\frac{D}{Dt}\frac{\delta \ell _1' }{\delta \mathbf{u}}=- \nabla\boldsymbol\nu\cdot\frac{\delta \ell _1'}{\delta \boldsymbol{\nu}}- \nabla\mathbf{n}\cdot\frac{\delta \ell_1}{\delta \mathbf{n}}-\nabla p, 
\quad \operatorname{div}\mathbf{u}=0,
\\
\vspace{0.2cm}\displaystyle
\frac{D}{D t}\frac{\delta
\ell_1'}{\delta\boldsymbol\nu}-\boldsymbol\nu\times\frac{\delta \ell_1}{\delta\boldsymbol\nu}=\mathbf{n}\times\frac{\delta
\ell_1'}{\delta \mathbf{n}},
\quad 
\frac{D\mathbf{n}}{Dt}=\boldsymbol\nu\times\mathbf{n}.
\end{array}\right.
\end{equation}
Upon inserting the variational derivatives
\[
\frac{\delta \ell'_1}{\delta \mathbf{u}}=\mathbf{u}
\,,\qquad
\frac{\delta \ell'_1}{\delta \boldsymbol\nu}=-J\,\mathbf{n}\times(\mathbf{n}\times \boldsymbol{\nu})
\,,\qquad
\frac{\delta \ell'_1}{\delta \bn}=-\mathbf{h}-J\,\boldsymbol{\nu}\times(\boldsymbol{\nu}\times\mathbf{n})
\]
and by repeating analogous steps to those in \S\ref{FromErToEL},
one obtains the hydrodynamic Ericksen-Leslie equations
\begin{equation}\label{Erick-Lesl}
\left\{
\begin{array}{l}
\vspace{0.2cm}\displaystyle
\frac{\partial\mathbf{u}}{\partial t}+(\mathbf{u}\cdot\nabla)\mathbf{u}=-\nabla p-\partial_i\!\left(\nabla\mathbf{n}\cdot\frac{\partial F}{\partial\mathbf{n}_{,i}}\right),
\quad \operatorname{div}\mathbf{u}=0,
\\
\displaystyle J\frac{D^2\mathbf{n} }{Dt^2}-2\left(\mathbf{n\cdot h}+J\,\mathbf{n}\cdot\frac{D^2\mathbf{n} }{Dt^2}\right)\mathbf{n}+\mathbf{h}=0\,.
\end{array}
\right.
\end{equation}
On the other hand, the Lagrangian \eqref{EringenFluidLagrangian} also gives the equations \eqref{EP_fluid2} in the form
\begin{equation}
\label{inc_Eringen}
\left\{
\begin{array}{l}
\vspace{0.2cm}\displaystyle
\frac{\partial\mathbf{u}}{\partial t}+(\mathbf{u}\cdot\nabla)\mathbf{u}=-\nabla p-\frac{\partial}{\partial x^i}\!\left(\frac{\partial \Psi}{\partial\gamma_i^a}\,\bgamma^a\right), \quad \operatorname{div}\mathbf{u}=0, \\
\displaystyle j\frac{D\bnu}{D t}=j\bnu\times\bnu-\frac{\partial}{\partial x^i}\frac{\partial \Psi}{\partial\bgamma_i}+{\boldsymbol\gamma^a}  \times\frac{\partial \Psi}{\partial\bgamma^a}\\
\vspace{0.2cm}\displaystyle  \frac{Dj}{D t} +[j,\hat{\boldsymbol{\nu}}]=0,\\
\displaystyle \partial _t \bgamma +(\mathbf{u}\cdot\nabla)\bgamma+\bgamma_i\nabla{u}^i+  \bgamma \times \boldsymbol{\nu}+ {\nabla} \boldsymbol{\nu}=0,\quad{\bgamma}_0=0\,,
\end{array}
\right.
\end{equation}
where the index $i$ refers to the spatial coordinates in 
$\mathcal{D}$ while the indexes $a,b,\dots$ refer to vectors 
in $\mathbb{R}^3$.

\begin{theorem}
Under the assumption $j = J(\boldsymbol{1} - 
\mathbf{n} \otimes \mathbf{n})$, Eringen's incompressible micropolar
system \eqref{inc_Eringen} with 
$\Psi(j, \boldsymbol{\gamma})$ given in 
\eqref{FrankMicropolar}, are equivalent to the 
Ericksen-Leslie liquid crystal equations \eqref{Erick-Lesl}.
\end{theorem} 

\paragraph{Proof.} Using relation \eqref{objectivity}, the
system of equations \eqref{inc_Eringen} becomes
\[
\left\{
\begin{array}{l}
\vspace{0.2cm}\displaystyle
\frac{\partial\mathbf{u}}{\partial t}+(\mathbf{u}\cdot\nabla)\mathbf{u}=-\nabla p-\frac{\partial}{\partial x^i}\!\left(\frac{\partial \Psi}{\partial\gamma_i^a}\,\bgamma^a\right), \quad \operatorname{div}\mathbf{u}=0, \\
\displaystyle j\frac{D\bnu}{D t}=j\bnu\times\bnu-
\operatorname{div}^{\boldsymbol{\gamma}} 
\frac{\partial\Psi}{ \partial \boldsymbol{\gamma}} + 
\overrightarrow{ \left[j,\frac{\partial \Psi}{ \partial j}
\right]}\\
\vspace{0.2cm}\displaystyle  \frac{Dj}{D t} +[j,\hat{\boldsymbol{\nu}}]=0,\\
\displaystyle \partial _t \bgamma +(\mathbf{u}\cdot\nabla)\bgamma+\bgamma_i\nabla{u}^i+  \bgamma \times \boldsymbol{\nu}+ {\nabla} \boldsymbol{\nu}=0,\quad{\bgamma}_0=0\,.
\end{array}
\right.
\]
The assumption $j = J(\mathbf{I} - 
\mathbf{n}\otimes\mathbf{n})$ takes this system to
\begin{equation}
\label{inc_Eringen_modified}
\left\{
\begin{array}{l}
\vspace{0.2cm}\displaystyle
\frac{\partial\mathbf{u}}{\partial t}+(\mathbf{u}\cdot\nabla)\mathbf{u}=-\nabla p-\frac{\partial}{\partial x^i}\!\left(\frac{\partial \Phi}{\partial\gamma_i^a}\,\bgamma^a\right), \quad \operatorname{div}\mathbf{u}=0, \\
\displaystyle 
J \mathbf{n} \times \left( \mathbf{n}\times \frac{D\bnu}{Dt}
\right)=
J\left(\mathbf{n} \times \left( \mathbf{n}\times \bnu\right)\right)\times\bnu+ \operatorname{div}^{\boldsymbol{\gamma}}
\frac{\partial \Phi}{\partial\bgamma} + 
\mathbf{n} \times\frac{\partial \Phi}{ \partial \mathbf{n}}\\
\vspace{0.2cm}\displaystyle  
\frac{D \mathbf{n}}{D t} +\mathbf{n} \times \boldsymbol{\nu}=0,\\
\displaystyle \partial _t \bgamma +(\mathbf{u}\cdot\nabla)\bgamma+\bgamma_i\nabla{u}^i+  \bgamma \times \boldsymbol{\nu}+ {\nabla} \boldsymbol{\nu}=0,\quad{\bgamma}_0=0\,,
\end{array}
\right.
\end{equation} 
where $\Phi(\mathbf{n}, \boldsymbol{\gamma})$ has been obtained from \eqref{FrankMicropolar} through the relation
\[
\Phi(\mathbf{n}, \boldsymbol{\gamma}) = \Psi(J(\mathbf{I} - \mathbf{n}\otimes \mathbf{n}), \boldsymbol{\gamma})
\]
and we have used the identity
\[
\mathbf{n} \times \frac{\partial \Phi}{ \partial\mathbf{n}}
= - \overrightarrow{
\left[j, \frac{\partial \Psi}{\partial j} \right]}\,.
\]
In turn, from \eqref{all_potential_relations}, we have 
$\Phi(\mathbf{n}, \boldsymbol{\gamma}) =  
F( \mathbf{n},\mathbf{n}\times \boldsymbol{\gamma}) = 
F(\mathbf{n}, \nabla\mathbf{n})$ as 
in \eqref{Frank-variant}.
Moreover, the relation $\nabla\mathbf{n} = 
\mathbf{n} \times \boldsymbol{\gamma}$, is used to give the
functional relations
\[
\frac{\partial \Phi}{ \partial \boldsymbol{\gamma}_i} = 
\frac{\partial F}{\partial \mathbf{n}_{,i}} \times \mathbf{n}.
\]  
Then, we conclude  that 
\[
\frac{\partial \Phi}{\partial\gamma_i^a}\,\bgamma^a = 
\nabla\mathbf{n}\cdot
\frac{\partial F}{\partial\mathbf{n}_{,i}}\,.
\]
Recalling relation \eqref{identity_useful_EL}, i.e.,
\[
\mathbf{n} \times  \mathbf{h}
=\operatorname{div}^{\boldsymbol{\gamma}}\frac{\partial \Phi}{ \partial\boldsymbol{\gamma}} + \mathbf{n} \times \frac{\partial \Phi}{\partial \mathbf{n} }\,,
\]
the second equation in \eqref{inc_Eringen_modified} becomes
\[
J \mathbf{n} \times \left( \mathbf{n}\times \frac{D\bnu}{Dt}
\right)=
J\left(\mathbf{n} \times \left( \mathbf{n}\times \bnu\right)\right)\times\bnu+ \mathbf{n} \times\mathbf{h}
\]
which can be rewritten as
\[
\frac{D}{ Dt}\left( \mathbf{n} \times (\mathbf{n} \times \boldsymbol{\nu})\right) =\mathbf{n} \times \mathbf{h}
\]
upon using the Jacobi identity. Recalling the third equation
in system \eqref{inc_Eringen_modified}, and taking the cross
product on the left with $\mathbf{n}$ gives second equation in 
\eqref{Erick-Lesl}. Therefore, the 
$\boldsymbol{\gamma}$-equation
decouples in \eqref{inc_Eringen_modified} which proves the
theorem. 
\quad $\blacksquare$

\begin{remark}[One constant approximation with fluid flow]
{\rm Using the same argument as in Remark \ref{rem:one_constant}, we may easily specialize equations 
\eqref{inc_Eringen} to the case of the one constant approximation of the free energy, i.e., 
$\Psi(j,\bgamma)=\frac{K}{2J}(\bgamma_i\cdot j\bgamma_i)$.
Indeed, the last summand in the first equation of system
\eqref{inc_Eringen} becomes
\[
\frac{\partial}{\partial x^i}\!\left(\frac{\partial \Psi}{\partial\gamma_i^a}\,\bgamma^a\right) = \frac{K}{J}
\frac{\partial}{\partial x^i}\!\left(j^{ab}\gamma^b_i\,\bgamma^a\right)  = \frac{K}{J}\boldsymbol{\gamma}^a \cdot 
\left((j \boldsymbol{\gamma}_i) \times \boldsymbol{\gamma}_i\right)^a +\frac{K}{J}j^{ab} \frac{\partial}{\partial x^i}\!\left(
\gamma_i^b \boldsymbol{\gamma}^a\right)
\]
by using  the relation ${\partial}_{x_i} j = \left[j, \widehat{\boldsymbol{\gamma}}_i\right]$. We get
\begin{equation*}
\left\{
\begin{array}{l}
\vspace{0.2cm}\displaystyle
\frac{\partial\mathbf{u}}{\partial t}+(\mathbf{u}\cdot\nabla)\mathbf{u}=-\nabla p- \frac{K}{J} \left(\boldsymbol{\gamma}^a \cdot 
\left((j \boldsymbol{\gamma}_i) \times \boldsymbol{\gamma}_i\right)^a +j^{ab} \frac{\partial}{\partial x^i}\!\left(
\gamma_i^b \boldsymbol{\gamma}^a\right)
\right), \quad \operatorname{div}\mathbf{u}=0, \\
\vspace{0.2cm}
\displaystyle J\big(j\partial_t{\bnu} + j(\mathbf{u}\cdot \nabla )\boldsymbol{\nu} -
(j\bnu)\times\bnu\big)=-K\big(j{\partial_{x_i}\bgamma_i}
+ (j\bgamma_i)\times\bgamma_i \big), \\
\vspace{0.2cm}\displaystyle \partial_t j+ 
(\mathbf{u} \cdot\nabla)j 
+[j,\hat{\boldsymbol{\nu}}]=0,\\
\displaystyle \partial _t \bgamma_k +
(\mathbf{u}\cdot \nabla)\boldsymbol{\gamma}_k 
+ \boldsymbol{\gamma}_i {\partial_{x_k}}u^i 
+ \bgamma_k \times \boldsymbol{\nu}+ {\partial_{x_k}} \boldsymbol{\nu}=0,\qquad{\bgamma}_0=0\,.
\end{array}
\right. 
\end{equation*}
The above equations can also be expressed in terms of the director field $\bn$, upon using the relation \eqref{microtensor}. It is interesting to observe that the equations above can be rewritten in the  form
\begin{equation*}
\left\{
\begin{array}{l}
\vspace{0.2cm}\displaystyle
\frac{\partial\mathbf{u}}{\partial t}+(\mathbf{u}\cdot\nabla)\mathbf{u}=-\nabla p- \frac{K}{J} \left(\boldsymbol{\gamma}^a \cdot 
\left((j \boldsymbol{\gamma}_i) \times \boldsymbol{\gamma}_i\right)^a +j^{ab} \frac{\partial}{\partial x^i}\!\left(
\gamma_i^b \boldsymbol{\gamma}^a\right)
\right), \quad \operatorname{div}\mathbf{u}=0, \\
\vspace{0.2cm}
\displaystyle J(\partial_t+\mathbf{u}\cdot\nabla)\!\left({j\bnu}\right)+K\partial_{x_i\!}\!\left({j\bgamma_i}\right)=0, \\
\vspace{0.2cm}\displaystyle \partial_t j+ 
(\mathbf{u} \cdot\nabla)j 
+[j,\hat{\boldsymbol{\nu}}]=0,\\
\displaystyle \partial _t(j \bgamma_k )+
(\mathbf{u}\cdot \nabla)(j\boldsymbol{\gamma}_k )
+(j \boldsymbol{\gamma}_i) {\partial_{x_k}}u^i 
+ \boldsymbol{\nu}\times j \bgamma_k + j{\partial_{x_k}} \boldsymbol{\nu}=0,\qquad{\bgamma}_0=0\,,
\end{array}
\right. 
\end{equation*}
which are then accompanied by  ${\partial}_{x_i} j = \left[j, \widehat{\boldsymbol{\gamma}}_i\right]$.
}\qquad $\blacklozenge$
\end{remark} 

\begin{remark}[Compressible flows]{\rm We have presented above the incompressible
case. Everything works out in the same way for the corresponding compressible variants. \quad $\blacklozenge$
}
\end{remark}

\section{Conclusions}

In this paper, we have proved the equivalence among various descriptions of conservative liquid crystal dynamics. This was achieved by applying reduction by symmetry in a systematic way, starting from the Hamilton's principle formulation for Ericksen-Leslie dynamics in the material description, whose Lagrangian is given in \eqref{UnreducedLagrangian}. 

By using Eringen's gauge-invariant definition of the wryness tensor $\bgamma=-(\nabla\chi)\chi^{-1}$, we used the resulting relation $\nabla\bn=\bn\times\bgamma$ to reformulate Ericksen-Leslie theory in terms of the wryness tensor dynamics. This has led to the new set of dynamic equations presented in \eqref{EL_Eringen-gamma}.

As a second step, we established the equivalence between  Eringen's micropolar theory in \eqref{EP3-B} and a new formulation of liquid crystal dynamics, whose order parameter field is identified with the microinertia tensor $j$. Indeed, the invariant relation $\nabla j=\left[j,\widehat{\gamma}\right]$ has led to the new set of dynamic equations presented in \eqref{EP3-A}. Then, we implemented the additional invariant relation $j=J(\mathbf{I}-\bn\otimes\bn)$ to recover the two equivalent theories found previously, whose equations of motion are \eqref{EL_Eringen-gamma} and the Ericksen-Leslie equations in the form \eqref{EP_form_EL}. As a consequence, 
we showed that the relations $\nabla j=
\left[j,\widehat{\gamma}\right]$ and 
$j=J(\mathbf{I}-\bn\otimes\bn)$ establish the equivalence of the four theories given by the systems \eqref{EP_form_EL}, \eqref{EL_Eringen-gamma}, \eqref{EP3-B}, and \eqref{EP3-A}.

Finally, we considered the model proposed by Lhuillier and Rey in \cite{LR}, which was shown to allow for an alternative formulation in terms of Eringen's wryness tensor. We have
used the relation $\nabla\bn=\bn\times\bgamma$ to transform the Lhuillier-Rey equations \eqref{EP_form_EL} into the new system \eqref{EL_Eringen-gamma}.

In conclusion, in this paper we established the equivalence of the six systems \eqref{EP_form_EL}, \eqref{EL_Eringen-gamma}, \eqref{EP3-A},  \eqref{EP3-B}, \eqref{EP_form_EL}, and \eqref{EL_Eringen-gamma}. In particular, we identified explicit necessary conditions for the equivalence of these six systems. These conditions were given in the theorems reported in the various sections and always require  a zero value for the initial wryness tensor $\bgamma_0$ as well as the rod-like assumption $j=J(\mathbf{I}-\bn\otimes\bn)$ on the microinertia tensor.
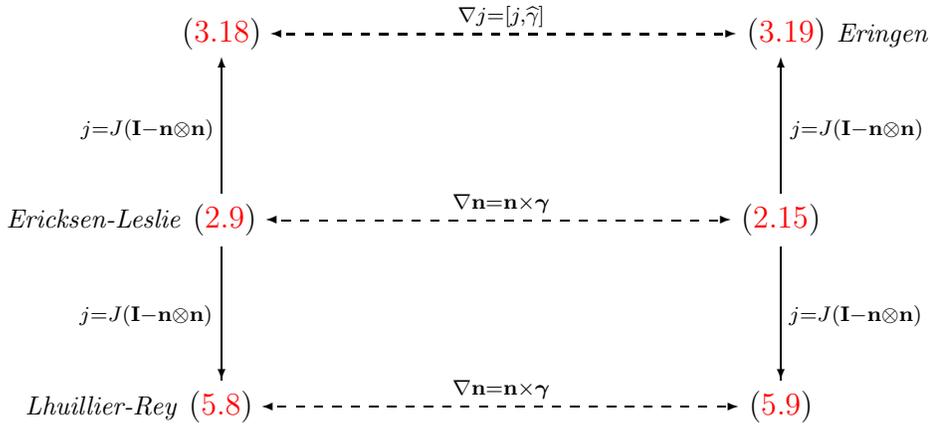
\begin{diagram}
\eqref{EP3-A}& &  & \lCorresponds{ \nabla j=\left[j,\widehat{\gamma}\right]} &  &&  \text{\eqref{EP3-B} {\it\footnotesize Eringen}}\hspace{-1.5cm}
\\
\uTo^{j=J(\mathbf{I}-\bn\otimes\bn)}&& & &&&\uTo_{j=J(\mathbf{I}-\bn\otimes\bn)}
\\
\hspace{-2.4cm}\text{{\it \footnotesize Ericksen-Leslie} \eqref{EP_form_EL}}& & & \rCorresponds^{\nabla \bn=\bn\times\bgamma} & & & \eqref{EL_Eringen-gamma}
\\
\dTo^{j=J(\mathbf{I}-\bn\otimes\bn)}&& &  &&&\dTo_{j=J(\mathbf{I}-\bn\otimes\bn)}
\\
\hspace{-2.2cm}\text{{\it \footnotesize Lhuillier-Rey} \eqref{EP-LR1-A}}& &  & \lCorresponds{ \nabla \bn=\bn\times\bgamma} &  &&  \eqref{EP-LR1-B}
\end{diagram}
This diagram relates the various equations of motion in each of the six models. The arrows characterize the transformations that are required to pass from one model to the other. 
In particular, the relations attached to the various arrows clarify the conversions that transfer one system to another, in the sense of the arrow. The dashed arrows emphasize the relations arising from the two different reductions on the same Lagrangian in each case.

Having equivalent equations of motion for 
the same model is advantageous since certain questions may 
be easier to treat in one of the formulations as opposed to 
the other. For example, in \cite{ChRaRoSa2013}, short time 
existence and uniqueness of strong solutions for the
initial value problem for the viscous non-dissipative EL equations (the dissipative part of the stress tensor and the dissipative part of intrinsic body force are set equal to zero) was proved in two 
situations: the space-periodic problem and the case of a 
bounded domain with spatial Dirichlet boundary conditions 
on the Eulerian velocity and the cross product of the 
director field with its time derivative. However, the paper 
did not work directly with the viscous non-dissipative 
version of the EL equations \eqref{EP_form_EL} on a two
dimensional domain, but with the equivalent viscous 
non-dissipative version of system \eqref{EL_Eringen-gamma}; an 
upper bound for the speed of propagation of the director 
field is also given in \cite{ChRaRoSa2013}.

\bigskip

Notice that, although this paper was restricted to consider conservative dynamics, the invariant relations $j=J(\mathbf{I}-\bn\otimes\bn)$ and $\nabla j=\left[j,\widehat{\gamma}\right]$ are not affected by the possible presence of dissipative effects that can be included within the various theories, for example by Rayleigh's method (see \cite{GBRaTr2012}). Moreover, we observe that the generality of the reduction processes guarantees that the methods apply to dynamical theories involving probability density functions encoding the microscopic properties of nematodynamics. See \cite{Tronci2012} for how these methods apply in this context within the Hamiltonian framework.

The results in this paper are summarized in the previous diagram, while the following one relates the various Lagrangians underlying the different theories.  
\bigskip

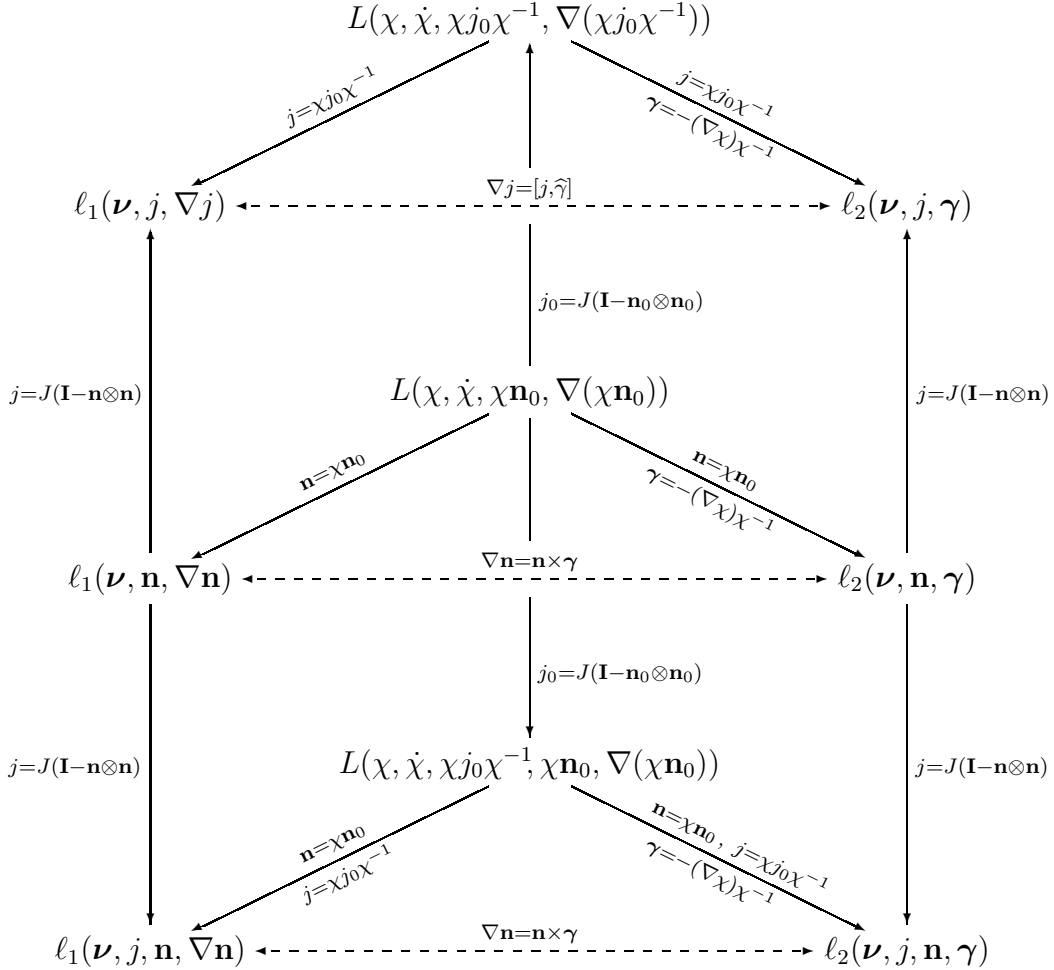
\[\hspace{-1.2cm}
\begin{diagram}
& &  & L(\chi,\dot\chi,\chi j_0 \chi^{-1},\nabla(\chi j_0 \chi^{-1})) &  && 
\\
&&\ldTo(3,2){j=\chi j_0\chi^{-1}}& \uTo&\rdTo(3,2)^{j=\chi j_0\chi^{-1}}_{\bgamma=-(\nabla\chi)\chi^{-1}}&&
\\
\ell_1(\bnu,j,\nabla j)& & & \lCorresponds{ \nabla j=\left[j,\widehat{\gamma}\right]}  & \HonV &&  \ell_2(\bnu,j,\bgamma)
\\
&&&\vLine_{j_0=J(\mathbf{I}-\bn_0\otimes\bn_0)}&&&
\\
\uTo^{j=J(\mathbf{I}-\bn\otimes\bn)}&& & L(\chi,\dot\chi,\chi \bn_0,\nabla(\chi \bn_0)) &&&\uTo_{j=J(\mathbf{I}-\bn\otimes\bn)}
\\
&&\ldTo(3,2)^{\bn=\chi\bn_0}&\vLine& \rdTo(3,2)^{\bn=\chi\bn_0}_{\bgamma=-(\nabla\chi)\chi^{-1}}&&
\\
\ell_1(\bnu,\bn,\nabla \bn)& & & \rCorresponds^{\nabla \bn=\bn\times\bgamma} & & & \ell_2(\bnu,\bn,\bgamma)
\\
&&&\dTo_{j_0=J(\mathbf{I}-\bn_0\otimes\bn_0)}&&&
\\
\dTo^{j=J(\mathbf{I}-\bn\otimes\bn)}&& & L(\chi,\dot\chi,\chi j_0\chi^{-1\!\!},\chi \bn_0,\nabla(\chi \bn_0)) &&&\dTo_{j=J(\mathbf{I}-\bn\otimes\bn)}
\\
&&\ldTo(3,2)^{\bn=\chi\bn_0}_{j=\chi j_0 \chi^{-1}}&& \rdTo(3,2)^{\quad \bn=\chi\bn_0\,,\  j=\chi j_0\chi^{-1}}_{\bgamma=-(\nabla\chi)\chi^{-1}}&&
\\
\ell_1(\bnu,j,\bn,\nabla \bn)& &  & \lCorresponds{ \nabla \bn=\bn\times\bgamma} &  &&  \ell_2(\bnu,j,\bn,\bgamma)
\end{diagram}
\]
\smallskip

\noindent
The Lagrangians on the center line identify the material descriptions of the models. The slanted arrows denote the Euler-Poincar\'e reduction processes while vertical arrows show how the theories are embedded in each other. Notice that the angular frequency $\bnu$ is defined everywhere by the relation $\widehat{\nu}:={\dot\chi}\chi^{-1}$. By Euler-Poincar\'e reduction theory, all the various Lagrangians related by a dashed arrow are equivalent. One of the
consequences of this diagram is that any concrete question
in a given model can be treated, equivalently, with any
of the three Lagrangians in a given triangle.


\appendix

\section{Expression of the Frank energy}\label{Appendix}

We show that the relation $\nabla\mathbf{n}= -\boldsymbol \gamma\times\mathbf{n}$ allows the Frank free energy $F(\mathbf{n},\nabla{\mathbf{n}})$ to  be expressed only in terms of $\boldsymbol\gamma(x)$ and $\mathbf{n}(x)$, so that $F(\mathbf{n}(x),\nabla{\mathbf{n}}(x))=\Psi(\mathbf{n}(x),\boldsymbol\gamma(x))$.

\paragraph{Twist.} We compute
\begin{align*}
\mathbf{n}\cdot(\nabla\times\mathbf{n})
&=- \frac{1}{2} \operatorname{Tr}\left(
\widehat{ \mathbf{n}}2 (\nabla \mathbf{n} )^A \right) 
=-  \operatorname{Tr}\left(\widehat{ \mathbf{n}}  \nabla \mathbf{n} \right) \\
&=- \operatorname{Tr}\left(\widehat{ \mathbf{n}} ( \mathbf{n} \times \boldsymbol{\gamma })\right)
=- \operatorname{Tr}\left(\widehat{ \mathbf{n}} 
\widehat{\mathbf{n}}\boldsymbol{\gamma }\right) \\
&=-\mathbf{n}\cdot\boldsymbol{\gamma}(\mathbf{n})
+\|\mathbf{n}\|^2\operatorname{Tr}(\boldsymbol{\gamma}). 
\end{align*}

\paragraph{Splay.}
Define the vector $ \underline{\boldsymbol{\gamma }}$ by the equality
\[
\underline{\boldsymbol{\gamma }} \cdot \mathbf{u} = \operatorname{Tr} ( \mathbf{u} \times \boldsymbol{\gamma}),\quad\text{i.e.,}\quad \underline{\boldsymbol{\gamma }}^a=\epsilon_{iab} \boldsymbol{\gamma}^b_i\quad\text{or}\quad \widehat{\left(\underline{\boldsymbol{\gamma}}\right)}=-2\gamma^A
\]
and we compute
\begin{align*}
\left(\operatorname{div}\mathbf{n}\right)^2&= \left(\operatorname{Tr} ( \nabla \mathbf{n} )\right)^2 = \left(\operatorname{Tr} ( \mathbf{n}\times \boldsymbol{\gamma } )\right)^2=( \underline{\boldsymbol{\gamma }} \cdot \mathbf{n} )(\underline{\boldsymbol{\gamma }} \cdot \mathbf{n}). 
\end{align*} 

\paragraph{Bend.}
For all $ \mathbf{u} \in \mathbb{R}  ^3 $, we have
\begin{align*}
(\mathbf{n} \times ( \nabla \times \mathbf{n} ) ) 
\cdot \mathbf{u} 
&=-\nabla _{ \mathbf{n} } \mathbf{n}  \cdot \mathbf{u} 
=- (\mathbf{n} \times \boldsymbol{\gamma}( \mathbf{n} )) \cdot \mathbf{u} =-(\mathbf{u} \times \mathbf{n} ) \cdot \boldsymbol{\gamma}( \mathbf{n} ) \\
&=- \widehat{\mathbf{u}} \mathbf{n}  \cdot 
\boldsymbol{\gamma}( \mathbf{n} ) 
= -\operatorname{Tr}((\widehat{ \mathbf{u} } \mathbf{n})^T \boldsymbol{\gamma}(\mathbf{n}))\nonumber \\
&= \operatorname{Tr} \left(  (\mathbf{n} \otimes 
\mathbf{n})\widehat{ \mathbf{u} } \boldsymbol{\gamma}\right),
\end{align*} 
so we get
\[
\mathbf{n} \times ( \nabla \times \mathbf{n} )=\underline{\boldsymbol{\gamma}(\mathbf{n}\otimes\mathbf{n})}
=-\mathbf{n}\times\boldsymbol{\gamma}(\mathbf{n})
\]
and therefore,
\[
\|\mathbf{n} \times ( \nabla \times \mathbf{n} )\|^2=\|\mathbf{n}\times\boldsymbol{\gamma}(\mathbf{n})\|^2=\|\mathbf{n}\|^2\|\boldsymbol{\gamma}(\mathbf{n})\|^2- (\mathbf{n}\cdot\boldsymbol{\gamma}(\mathbf{n}))^2
\]

Putting all these results together, we conclude that  there exists a function $\Psi(\mathbf{n}(x),\boldsymbol\gamma(x))$, such that $\Psi(\mathbf{n}(x),\boldsymbol\gamma(x))=F(\mathbf{n}(x),\nabla\mathbf{n}(x))$, that is,
\begin{align*}
\Psi(\mathbf{n}(x),\boldsymbol\gamma(x))
&=-K_2\left(\|\mathbf{n}\|^2\operatorname{Tr}
(\boldsymbol{\gamma})
-\mathbf{n}\cdot\boldsymbol{\gamma}(\mathbf{n})\right)
+\frac{1}{2}K_{11}( \underline{\boldsymbol{\gamma }} \cdot 
\mathbf{n} )^2\\
&\qquad\qquad+\frac{1}{2}K_{22}\left(\|\mathbf{n}\|^2 
\operatorname{Tr}(\boldsymbol{\gamma})-\mathbf{n}\cdot
\boldsymbol{\gamma}(\mathbf{n})\right)^2\\
&\qquad\qquad +\frac{1}{2}K_{33}\left(\|\mathbf{n}\|^2\|\boldsymbol{\gamma}(\mathbf{n})\|^2- (\mathbf{n}\cdot\boldsymbol{\gamma}(\mathbf{n}))^2\right).
\end{align*}

{\footnotesize

\bibliographystyle{new}

}

\end{document}